\documentclass{aa}

\usepackage{graphicx,multirow,threeparttable,lscape}
\usepackage{txfonts}
\bibpunct{(}{)}{;}{a}{}{,}
\begin{document}

   \title{CI Camelopardalis: The first sgB[e]-High Mass X-ray Binary Twenty Years on}

    \subtitle{A Supernova Imposter in our own Galaxy?}
   
   \author{E. S. Bartlett\inst{1} \and J. S. Clark\inst{2} \and I. Negueruela\inst{3}}
       
   \institute{ESO - European Southern Observatory, Alonso de C{\' o}rdova 3107, Vitacura, Casilla 19001, Santiago de Chile, Chile\\\email{ebartlet@eso.org}
   		\and
		Department of Physics and Astronomy, The Open University, Walton Hall, Milton Keynes, MK7 6AA, UK
		\and
		Departamento de F\'{\i}sica Aplicada, Facultad de Ciencias, Universidad de Alicante, Carretera San Vicente del Raspeig s/n,\\
E03690, San Vicente del Raspeig, Spain\\
		}

   \date{Received September 25, 2018; accepted December 14, 2018}

 \abstract{The Galactic supergiant B[e] star CI Camelopardalis (CI Cam) was the first sgB[e] star detected during an X-ray outburst. The star brightened to $\sim$2 Crab in the X-ray regime ($\sim5\times10^{-8}$ ergs~cm$^{-2}$~s$^{-1}$ in the 2-25~keV range) within hours before decaying to a quiescent level in less than 2 weeks, clearly indicative of binarity. Since the outburst of CI Cam, a number of sgB[e] stars have been identified as X-ray overluminous for a single star (i.e., $L_X > 10^{-7}~L_{bol}$). This small population has recently expanded to include two Ultra Luminous X-ray sources (ULX), Holmberg II X-1 and NGC300 ULX-1/supernova imposter SN2010da.}
 {Since the discovery of X-ray emission from CI Cam, there have been many developments in the field of massive binary evolution. In light of the recent inclusion of two ULXs in the population of X-ray bright sgB[e] stars, we revisit CI Cam to investigate its behaviour over several timescales and shed further light on the nature of the compact object in the system, its X-ray outburst in 1998 and the binary system parameters.}
 {We analyse archival \emph{XMM-Newton} EPIC-pn spectra and light curves along with new data from \emph{Swift} and \emph{NuSTAR}. We also present high-resolution (R$\sim$85,000) MERCATOR/HERMES optical spectra, including a spectrum taken 1.02 days after our \emph{NuSTAR} observation.}
 {Despite being in quiescence, CI Cam is highly X-ray variable on timescales of days, both in terms of total integrated flux and spectral shape. We interpret these variations by invoking the presence of an accreting compact companion immersed in a dense, highly structured, aspherical circumstellar envelope. The differences in the accretion flux and circumstellar extinction represent either changes in this environment, triggered by variable mass loss from the star, or the local conditions to the accretor due to its orbital motion. We find no evidence for pulsations in the X-ray light curve.}
 {CI Cam has many similarities with SN2010da across mid-IR, optical and X-ray wavelengths suggesting that, subject to distance determination for CI Cam, if CI Cam was located in an external galaxy its 1998 outburst would have led to a classification as a supernova imposter.}

   \keywords{Stars: emission-line, Stars: massive, Stars: individual: CI Camelopardalis, Stars: binaries, X-rays: binaries
               }

  \authorrunning{E. S. Bartlett et al.}
  \titlerunning{CI Cam - a SNe imposter in our own Galaxy?}
   \maketitle
   
%

   \section{Introduction}

The optical spectra of the evolved B[e] supergiants (sgB[e]s) are characterised by broad, high excitation emission lines; narrow, low excitation emission lines; and a strong infrared excess due to hot dust \citep{Lamers1998}. The standard explanation for this hybrid spectrum is a low density, fast polar wind and a much denser, slowly expanding equatorial outflow or torus in which dust forms \citep{Zickgraf1985,Zickgraf1986}. Recent observations suggest that the equatorial material of many sgB[e]s is in Keplerian rotation rather than radially expanding (e.g. \citealt{Kraus2010}) with \citet{Maravelias2018} demonstrating that each of their studied sgB[e] stars showed a unique pattern of concentric rings of material, from which the forbidden emission arises. The formative agent of the equatorial material is a topic of active research with stellar rotation suggested as one potential cause. The rotational velocities of a handful of sgB[e] star's have been determined, however, and only 1 of the 3 are rotating at an appreciable fraction of the critical velocity \citep{Gummersbach1995,Zickgraf2000,Zickgraf2006a,Kraus2008,Kraus2010}. There is mounting support for the idea that sgB[e] stars represent either interacting or newly formed, post- interaction binary systems. In this scenario, the tori are ejected during non-conservative mass transfer in an evolutionary stage of close binary evolution (e.g. the colliding wind binary Wd1-9, \citealt{Clark2013a}), or are material from the stellar wind captured into circum-binary orbits during periastron passages of the secondary (considered the most plausible scenario for the material surrounding GG\,Car, \citealt{Kraus2013}). At least one sgB[e] star, R4 in the Small Magellanic Cloud, is though to be the result of a binary merger \citep{Podsiadlowski2006}. Identifying bona fide sgB[e] binaries is a priority as the asymmetries seen in the circumstellar environment around the progenitor of SN1987A \citep{Morris2009} and SN2009ip \citep{Mauerhan2014} may be established in the sgB[e] phase. For a recent overview of sgB[e] stars, see \citealt{deWit2014}.

sgB[e] stars are co-located on the HR diagram with Luminous Blue Variables (LBVs). LBVs are thought to represent a short-lived phase of very massive stellar evolution in which instabilities drive dramatic mass loss \citep{Humphreys1994}. The physical origin of these instabilities remains poorly understood. Recent observational findings have associated a subset of highly luminous type II SNe with LBVs \citep{Gal-Yam2009} contradicting the classical view of these objects. An additional population of LBVs has been identified in external galaxies - the so-called SNe imposters - with significantly more rapid and violent eruptions (e.g. \citealt{Kochanek2012}). Unlike LBVs or SNe imposters, historical consensus has been that sgB[e] stars do not show significant variability ($\Delta m \lesssim 0.2$~mag;  \citealt{Lamers1998}), however, an evolutionary link between sgB[e] stars and LBVs has been suggested \citep{Zickgraf2006b} and some sgB[e] stars are considered candidate LBVs (e.g. \citealt{Clark2013a}). 

 A number of sgB[e] stars are X-ray over-luminous for single stars (i.e., $L_X > 10^{-7} L_{bol}$). Again, binarity is invoked to explain this with the excess high energy emission arising either via shocks in a colliding wind binary (CWB) containing a second massive star, or via accretion onto a compact companion. This list includes the Galactic sources Wd1-9, a CWB \citep{Clark2013b}; IGR~J16318-4848, thought to host a neutron star and thus be a bona fide High-Mass X-ray Binary (HMXB, e.g. \citealt{Filliatre2004}); and CI Camelopardalis \citep{Bartlett2013}, the subject of this paper. In the Magellanic Clouds, LHA 115-S 18 (S18) and LHA 120-S 134 (S134) are both reported to have X-ray emission (Bartlett in prep, \citealt{Clark2013a,Massey2014}), though their optical spectra display markedly different behaviour, suggesting they are fundamentally different types of systems \citep{Bartlett2015}. S134 is most likely a CWB while the nature of S18 is uncertain due to the unprecedented nature of both photometric and spectroscopic variability. \citet{Clark2013a} show that S18 trivially satisfies the criteria for an LBV, but its behaviour more closely resembles that of SNe imposters such as SN2000ch \citep{Pastorello2010}.
  
 Recently, two Ultra Luminous X-ray sources (ULX), Holmberg II X-1 and NGC\,300 ULX-1/supernova imposter SN2010da (referred to as SN2010da throughout this paper) have been interpreted as sgB[e]-HMXBs \citep{Lau2016,Lau2017,Villar2016}, with X-ray pulsations reported in SN2010da, clear evidence of a neutron star companion \citep{Carpano2018}. The association of two ULXs with sgB[e] stars highlights the need for further study into the link between binarity, sgB[e] stars, LBVs, SNe imposters and ULXs.  

   \subsection{CI Camelopardalis}

   The sgB[e] star CI Camelopardalis (CI Cam) underwent a dramatic X-ray outburst in1998. First detected 31 March \citep{Smith1998}, it increased in flux by a factor $>50$ within a day, to a peak flux of 5.4$\times10^{-8}$~ergs~cm$^{-2}$~s$^{-1}$, before decaying back down to 9.4$\times10^{-11}$~ergs~cm$^{-2}$~s$^{-1}$ in $\sim8$~days \citep{Belloni1999}. A \emph{BeppoSAX} observation, taken 3-4 September 1998 detected the source with a flux of 1.5$\times10^{-13}$~ergs~cm$^{-2}$~s$^{-1}$ \citep{Orlandini2000}: A drop of 5 orders of magnitude. The X-ray outburst was accompanied by increased emission at optical and radio wavelengths. In the optical, the $V$-band magnitude decreased by $>2$~mag and the source reddened significantly ($\Delta(B-V)\sim0.5$~mag, \citealt{Clark2000}). The radio emission peaked later than the optical (see Fig. 1. of \citealt{Clark2000}), and the radio images show an evolution from an unresolved source to a more complex geometry consistent with that expected for an expanding shock front \citep{Mioduszewski2004}.
   
Such X-ray activity firmly establishes this source as a HMXB but the nature of the compact object, as well as binary system parameters, still remain a mystery 20 years on. For example, \citet{Barsukova2006} and \citet{Filippova2008} favour a white dwarf companion and interpret the 1998 outburst was thermonuclear runaway on the white dwarf surface. \citet{Bartlett2013} make the case for a neutron star companion, based on the quiescent X-ray spectrum of CI Cam, but offer no mechanism for the outburst. \citet{Robinson2002} argue that the ratio of peak to quiescent X-ray luminosity is consistent with those of black hole X-ray novae. The distance to CI Cam is still not certain, with estimates ranging from 2.0--10 kpc in the literature but a distance of $\gtrsim8$~kpc implies a peak outburst luminosity of $10^{39}$~ergs~s$^{-1}$ \citep{Kaaret2017} i.e. CI Cam would be a ULX in our own galaxy (see \citealt{Belloni1999}). 

   In 2016, it was reported in an Astronomers Telegram (ATel) by \citeauthor{Wijngaarden2016} that CI Cam was undergoing another optical outburst, with the H$\alpha$ emission line comparable in strength to that measured in 1998. Motivated by this we requested, and were approved, both a snapshot Target of Opportunity (ToO) observation with The \emph{Nuclear Spectroscopic Telescope Array} \emph{NuSTAR} and a rolling medium term monitoring campaign with the \emph{Neil Gehrels Swift Observatory} (\emph{Swift}, initially reported in \citealt{Bartlett2016}). In this paper we report on the results of these ToO observations as well as revisiting both published and unpublished archival observation of CI Cam, taken with \emph{XMM-Newton}.
 
   {\footnotesize \begin{table*}
\centering
\caption{Log of the \emph{XMM-Newton} and \emph{NuSTAR}  snapshot observations of CI Cam.}\label{table:xmm_nustar}
\begin{tabular}{clccc}
\hline\hline\noalign{\smallskip}
\multirow{3}{*}{Detector}			 								&  \multirow{3}{*}{Obs. ID}			& \multirow{3}{*}{Obs. Date}		&\multicolumn{2}{c}{Exp. Time}	\\\noalign{\smallskip}
     															&							&							& unfiltered	& filtered 			\\\noalign{\smallskip}
    															&							&							&	(ks)		& (ks)				\\\noalign{\smallskip}
\hline\noalign{\smallskip}
\multirow{4}{*}{\parbox{2.1cm}{\centering \emph{XMM-Newton}: EPIC-pn}}	& 0000110101					& 2001-08-19					& 29.70		& 13.53	\\\noalign{\smallskip}
  															& 0139760101					& 2003-02-24					& 60.34		& 24.56	\\\noalign{\smallskip}
															& 0672050101					& 2012-02-20					& 67.40		& 32.64	\\\noalign{\smallskip}
    															& 0672050201					& 2012-02-22					& 68.64		& 23.82 	\\\noalign{\smallskip}
\emph{NuSTAR}: FMPA											& \multirow{2}{*}{90201040002}		& \multirow{2}{*}{2016-10-20}		& 20.00 		& 18.70	\\\noalign{\smallskip}
\emph{NuSTAR}: FMPB											&							&							& 20.01		& 18.65	\\\noalign{\smallskip}
      \hline
\end{tabular}
\end{table*}}


   \section{Observations and Data Acquisition}

   \subsection{\emph{XMM-Newton}}
     CI Cam has been observed on four separate occasions by \emph{XMM-Newton}. The data from the European Photon Imaging Cameras (EPIC) pn detector \citep{Struder2001} were processed with the \emph{XMM-Newton} Science Analysis Software (\textsf{xmmsas}) v15.0.0. Table \ref{table:xmm_nustar} shows a log of the EPIC-pn observations, along with the \emph{NuSTAR} ToO observation.
     
The EPIC-pn data of all the observations were processed with the \textsf{xmmsas} task \textsf{epproc}. The data were then screened for periods of high background by examining the 10.0--12.0~keV count rate of the entire detector. All observations were heavily affected by periods of high background. The light curves were inspected by eye and epochs where the count rate exceeded 0.8 counts s$^{-1}$ in observation 0000110101 and 1.0 count s$^{-1}$ in observations 0139760101, 0672050101 and 0672050201 were removed. 

Light curves and spectra of CI Cam were extracted from the filtered datasets. The extraction region size was determined using the \textsf{xmmsas eregionanalyse} task, which uses images, exposure maps and background maps to calculate the optimum radius for source extraction by maximising the signal to noise. The resulting source extraction radii range from 16--29 arcseconds depending on the source counts. Several regions with 60 arcsecond radii, located on the neighbouring chips to CI Cam, were identified as potential background regions, and compared to see if they were statistically identical. One of these regions was then selected as the final background region for each of the observations.

For the light curves, ``single'' and ``double'' pixel events (\textsf{PATTERN $\leq$ 4}) were selected in the energy range 0.2--12.0~keV with event attribute flag \textsf{\#XMMEA\_EP}. A barycentric correction was applied using the \textsf{barycen} task, converting the photon arrival times from the default satellite reference frame to barycentric dynamical time. The background subtraction was performed using the \textsf{epiclccorr} task which not only corrects for the different extraction region radii used for the source and background regions, but also corrects for bad pixels, vignetting and quantum efficiency. The source and background spectra were extracted using the same regions as the light curves. A more conservative selection flag of \textsf{FLAG=0} was used to exclude events close to CCD gaps and bad pixels. The area of source and background regions were calculated with the \textsf{backscal} task and response matrix files were created using \textsf{rmfgen} and \textsf{arfgen}. The spectra and associated response files for each observation were grouped to have at least 25 counts per bin using the \textsf{HEASOFT} task \textsf{grppha}, and fit with \textsf{XSPEC} v12.9.0.

      \subsection{\emph{NuSTAR}}
      
     In response to the ATel of \citeauthor{Wijngaarden2016}, we were approved a 20~ks ToO observation with the \emph{NuSTAR} observatory \citep{Harrison2013}. The data from both of the Focal Plane Modules (FMPA and FMPB) were processed with the NuSTAR Data Analysis Software (\textsf{nustardas}) v1.6.1. The data were first processed with the \textsf{nustardas} task \textsf{nupipeline} to produce cleaned and calibrated event files. The details of the observations and the pre and post-filtering exposure times are included in Table \ref{table:xmm_nustar}. Spectra, response matrix files and light curves were then generated using the \textsf{nuproducts} task. Source light curves and spectra were extracted from a circular region with 50 arcsecond radius centred on CI Cam. Background light curves and spectra were extracted from a nearby, source free circular region with a 75 arcsecond radius. Again, the spectra and associated response files were grouped to have at least 25 counts per bin using the \textsf{HEASOFT} task \textsf{grppha}, and fit with \textsf{XSPEC} v12.9.0.

     \subsection{\emph{Swift}}
     
     Since the 1998 outburst, all the X-ray observations of CI~Cam have been single snapshots, with no short to medium term monitoring to characterise its quiescent X-ray state. We requested, and were approved, several target of opportunity programs with the \emph{Swift} X-ray Telescope (XRT), each consisting of 10$\times \sim1$~ks observations with a cadence of $\sim3$~days. All the observations were carried out in Photon Counting (PC) mode. The data were processed with the \emph{swift-XRT} data products generator\footnote{http://www.swift.ac.uk/user\_objects/} \citep{Evans2007, Evans2009}, which makes use of \textsf{HEASOFT v6.22}.
     
      \subsection{HERMES Mercator}
      
      CI Cam was observed with the High-Efficiency and high-Resolution Mercator Echelle Spectrograph (HERMES, \citealt{Raskin2011}) at the 1.2~m Mercator telescope at the Roque de los Muchachos Observatory on La Palma, for 1800~s on the nights of 2016-10-19 and 2016-10-23; bookending our \emph{NuSTAR} ToO. HERMES is a fibre-fed prism-cross-dispersed echelle spectrograph, equipped with an E2V 42-90 detector of 2048$\times$4608, 13.5$\mu$m pixels. The instrument was operated in the high resolution mode which results in a resolution of $R\sim85,000$ over the 377-900~nm wavelength range. Wavelength calibration is achieved by observing both thorium-argon and neon arc lamp spectra, taken at the beginning and end of the night. The resulting spectra have a radial velocity accuracy of $\sim50$~m~s$^{-1}$. This error is driven by variable environmental conditions, most notably fluctuations of atmospheric pressure \citep{Raskin2011}. This corresponds to a wavelength accuracy of 0.001~\AA{} at 6500~\AA{}. The resulting spectra have an unbinned signal to noise ratio of $\sim$1 in the continuum.


\section{Results}
\subsection{Optical Spectra}

Figure
\begin{figure*}
\centering
\includegraphics[width=1.0\textwidth]{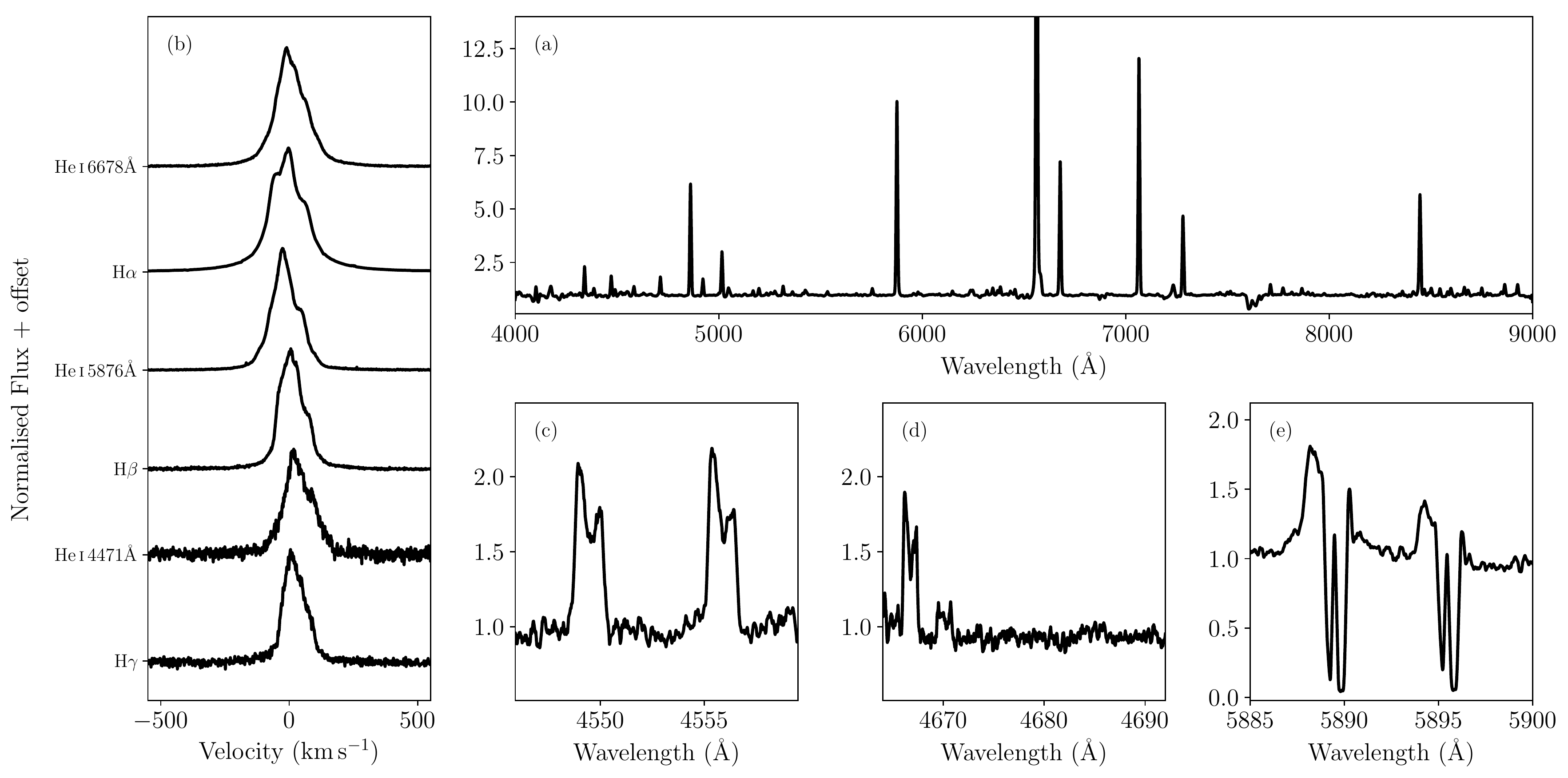}
\caption{The optical spectrum of CI~Cam. \emph{(a}) The 4000--9000~\AA{} spectrum of CI~Cam, convolved with a Gaussian function to approximately match the resolution of those presented in Fig. 1 of \citet{Hynes2002} to allow for easy comparison. \emph{(b)} The original, un-binned line profiles of the stronger hydrogen and helium lines in velocity space, corrected for the systemic velocity of CI Cam, -51$\pm2$~km~s$^{-1}$ \citep{Aret2016}. \emph{(c)} The systemic redshift corrected 4545--4560 \AA{} region, convolved with a boxcar function with width 13 pixels, corresponding 0.2~\AA{}. \emph{(d)} The region around the He \textsc{ii}~4686 \AA{} line, also redshift corrected and convolved with a boxcar function of width 13 pixels. \emph{(e)} The region around the Na\,{\sc D} interstellar absorption components, convolved with a boxcar function and with no redshift correction applied. }\label{fig:optical_spectrum}
\end{figure*} \ref{fig:optical_spectrum} shows the signal to noise weighted average of the two HERMES spectra of CI Cam, taken 2016-10-19 and 2016-10-23. No evidence of any variability between the two spectra was observed. Our spectra are consistent with the quiescent spectra of CI Cam \citep{Hynes2002,Robinson2002} with the helium lines similar in strength to those seen before the 1998 outburst (see Fig. 1. of \citealt{Hynes2002}). The Balmer series and strong He\textsc{i} lines are shown in subplot \emph{(b)} of Fig. \ref{fig:optical_spectrum}. Whilst the overall widths of our lines are comparable to the post outburst profiles of H$\alpha$ and He\,{\sc i} presented in Fig. 7. of \citet{Hynes2002} with Full Width Half Maximum values ranging from $\sim120$ to $\sim150$~km~s$^{-1}$, the line profiles appear much more complex, with up to three components visible in H$\beta$, H$\alpha$, He\,{\sc i} 5876~\AA{} and 6678~\AA{} lines. The He\,{\sc i} spectra of CI Cam in outburst (and just after, see Fig. 7 of \citealt{Hynes2002}) were well described with a narrow central component and a broader blue shifted line, these spectra suggest that such a model can be extended to the quiescent spectrum, along with an additional red shifted component.

The He\,{\sc ii}~4686\AA{}, which was strongly in emission during the 1998 outburst, and reported as weakly in emission in 2012 \citep{Goranskij2017} is completely absent here (see Fig \ref{fig:optical_spectrum}\emph{(d)}). In this regard CI Cam closely resembles the sgB[e]/LBV S18 \citep{Clark2013a}, although He\,{\sc ii} 4686{\AA} is, on occasion, considerably stronger in the latter system.  Another obvious comparator is the primary in SN2010da and once again both systems bear a close resemblance \citep{Villar2016}, with emission from both low (Fe\,{\sc ii}) and high excitation (He\,{\sc ii} and [O\,{\sc iii}]) species although SN2010da also demonstrates emission in very high excitation transitions (e.g. [Fe\,{\sc vii}] 5159, 5276 and 6086{\AA}) in outburst, that are absent in CI Cam at any epoch; we expand upon this comparison in Sect. \ref{sect:disc_comp}. We also highlight that the extreme variability of the He\textsc{ii} profile in both CI Cam and S18, both in terms of profile and overall strength, means that it forms an unreliable RV diagnostic. Similar conclusions were arrived at by \citet{Sana2012} who recommended against employing it for such a purpose (this is discussed further in Sect. \ref{sect:disc_obs}).

The metal lines, shown in Fig. \ref{fig:optical_spectrum}\emph{(c)} and \emph{(d)}, show the same concave shape reported in quiescence by \citet{Hynes2002}. The central depression in these lines is due to less emission at lower velocities, i.e. less emission from material in the equatorial plane of a spherically symmetric outflow around the star, or a more equatorial outflow viewed pole on (as is the case here). This is well demonstrated in Fig. 13 of \citet{Hynes2002}. A clear $V/R$ asymmetry is present with the blue peak stronger than the red. This behaviour is also suggested in the spectra presented by \citeauthor{Goranskij2017} (2017, see Fig 4.a) and could be indicative of a Global One-Armed Oscillation (GOAO) in the equatorial material, where these lines are hypothesised to originate \citep{Kato1983,Okazaki1991,Okazaki1997}. GOAOs are azimuthal density waves in with $m=1$ (the only mode permitted in a thin Keplerian disc, \citealt{Okazaki1991,Okazaki1997}) that can precess over time and are the accepted cause of $V/R$ variability in normal Be stars.

 \begin{figure*}
  \centering
  \includegraphics[width=0.49\textwidth]{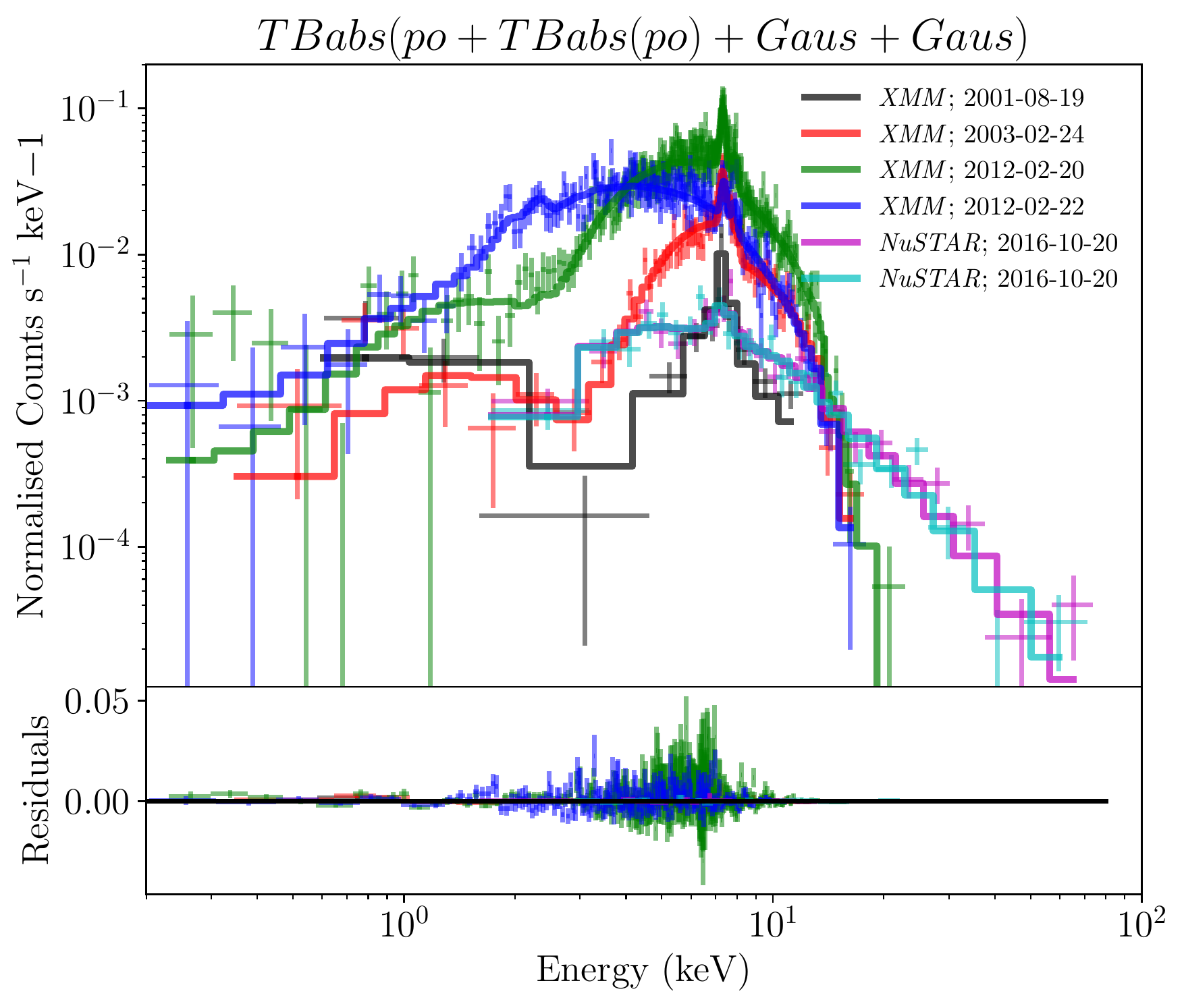}  \includegraphics[width=0.49\textwidth]{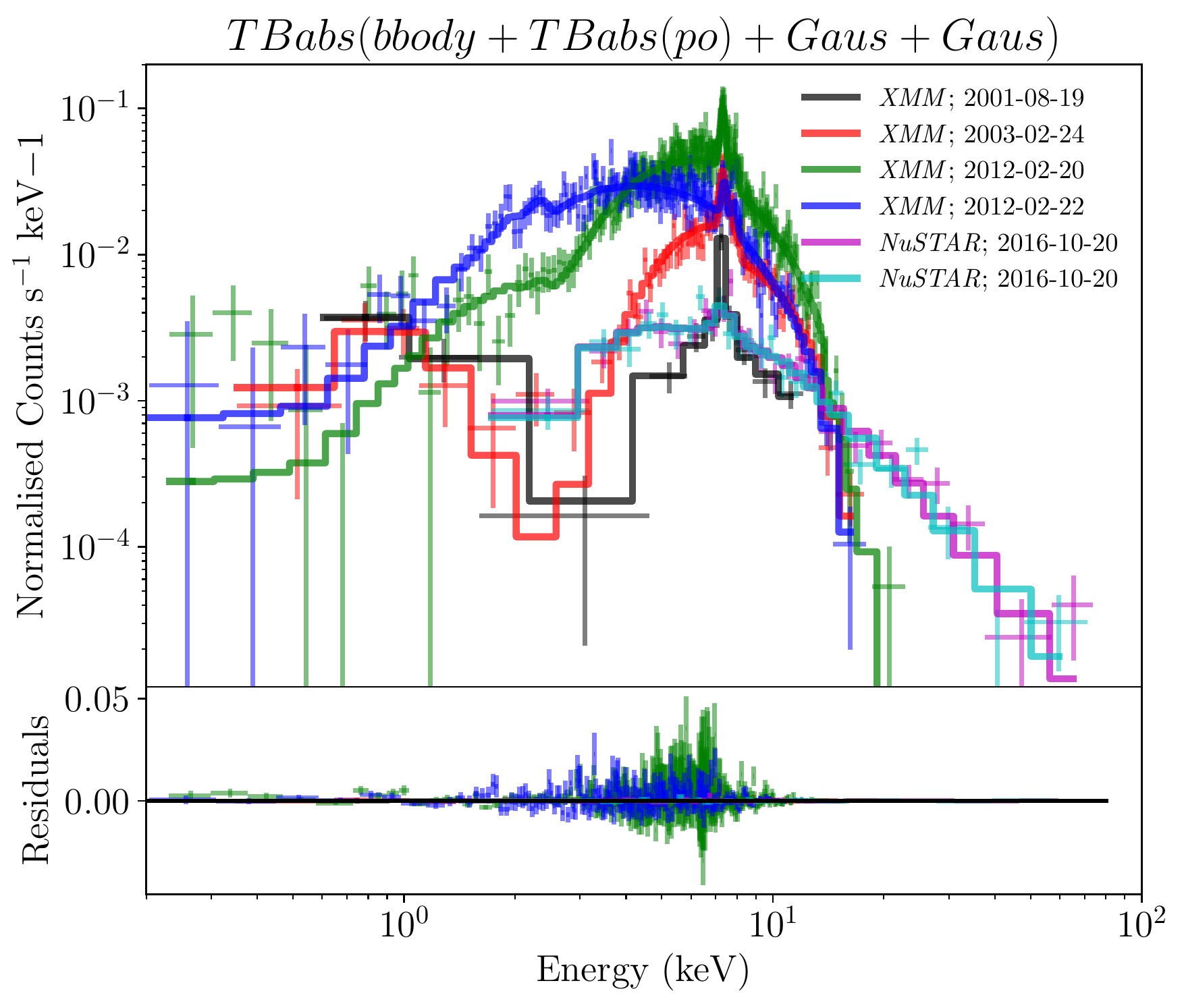}
  \caption{\emph{left:} \emph{XMM-Newton} and \emph{NuSTAR} data with the best fit \emph{TBabs(po+TBabs(po)+Gaus+Gaus)} model. \emph{right:} The same as the left but with the best fit \emph{TBabs(bbody+TBabs(po)+Gaus+Gaus)} model.}\label{fig:spectra}
\end{figure*}

\subsection{X-ray Spectra}\label{sect:xray_spec}

Figure \ref{fig:spectra} shows the archival spectra from \emph{XMM-Newton}, along with the recent \emph{NuSTAR} observations. In a previous study on the 2003 spectrum of this source, \citet{Bartlett2013} ignored the X-ray spectrum $<3$~keV, arguing that the soft excess in the 2003 spectra of CI Cam was likely  an artefact of the background subtraction. It is clear from the subsequent observations that not only is this is not the case, the $<3$~keV spectrum is also highly variable on timescales of days.

The spectra from 2001, 2003, and to a lesser extent the first spectrum of 2012, all resemble the 2010 spectrum of NGC 300 ULX1: a pronounced soft excess at ~< 2 keV, a minimum at around 3 keV and a strong emission feature at 6-7~keV (i.e. Fe-K emission, see Fig. 3 of \citealt{Carpano2018}). These authors fit an absorbed, partially covered disk blackbody and power law with a high energy cut off (i.e. \emph{TBabs*pcfabs(diskbb+po*highecut)} in \textsf{xspec}) model to the data. In their model, the photon index, cut off and e-folding energy, and disk blackbody temperature and flux are fixed across all the spectra, spanning 6 years, with just the intrinsic X-ray luminosity (i.e. the normalisation) and the absorption allowed to vary. Similar models have been used to fit the heavily obscured HMXB population discovered by \emph{INTEGRAL} as demonstrated by \citet{Fornasini2017}, who show that the spectrum of IGR~J18214-1318 is equally well fit by an absorbed, partially covered power law with a high energy cut off model (\emph{TBabs*pcfabs(po*highecut)} in \textsf{xspec}) or an absorbed blackbody and power law with a high energy cut off model (\emph{TBabs*(bbody+po*highecut)} in \textsf{xspec}).

Another possible comparator is the Supergiant Fast X-ray Transient (SFXT) IGR~J18410-0535 \citep{Bozzo2011}. This source was serendipitously observed in outburst during a scheduled 45~ks \emph{XMM-Newton} observation. Towards the end of the flare's decay the X-ray spectrum of IGR~J18410-0535 also showed the soft excess and iron line complex characteristic of the spectra of CI Cam. The authors fit the continuum of this spectrum with a model made up of two absorbed power laws (\emph{phabs*(po+phabs(po)} in \textsf{xspec}). The photon indices are constrained to be equal and the first absorption component is fixed to the Galactic value. The authors note that this model is often adopted to fit the X-ray spectra of ``normal'' supergiant X-ray Binaries in eclipse (e.g. \citealt{VanDerMeer2005}), and has also been used to fit at least one eclipsing Be/X-ray binary \citep{Coe2015}. In this scenario the more heavily absorbed power-law is the intrinsic emission from accretion onto the neutron star and the minimally absorbed power-law represents the scattered/reprocessed emission from the stellar wind material.

\subsubsection{The X-ray Continuum}

We adopt a similar technique as \citet{Bartlett2013} and first exclude the 6.2--7.0~keV region around the iron line and attempt to simultaneously fit the $>3$~keV continuum of all the EPIC-pn and \emph{NuSTAR} spectra with a simple absorbed power law model (\emph{TBabs*TBabs(po)} in \textsf{xspec}). The excluded region can be seen in Figure \ref{fig:Fe}. We intentionally included the Fe-K edge at 7.1~keV in our continuum fits. We then include $<3$~keV and add model components to reproduce and compare the continuum models discussed above, without the iron lines skewing the statistics. Finally, we re-introduce the iron line complex and fit the emission lines. As the data from both the \emph{NuSTAR} FPMA and FPMB detectors were taken simultaneously, all the model parameters of all the fits to these data were required to be the same, with a single constant factor to account for instrumental differences.

For the initial fit, the photon index of the power law, $\Gamma$, is required to be the same across all the spectra with a variable normalisation component to account for the different intrinsic luminosities of CI Cam. The absorption component is split into two parts $n_{H,Gal}$ to account for the Galactic interstellar absorption to CI Cam, fixed at $4.5\times10^{21}\text{~cm}^{-2}$ \citep{Dickey1990} and a variable component $n_{H,i}$, to account for absorption intrinsic to the system. This produces an acceptable fit to the data with a reduced $\chi^2$ ($\chi_r^2$) of 1.05 for 435 degrees of freedom (dof, $\chi^2= 457.4$), falling to 1.03 for 431 dof ($\chi^2=443.2$) when the photon indices are allowed to vary. The photon indices range from 1.1$\pm$0.2 from the 2003 spectra to 1.6$_{-0.1}^{+0.2}$ in that of 2016.

When the 0.2--3.0~keV energy range is included, the fit worsens dramatically ($\chi_r^2$=1.53 for 529 dof, $\chi^2$=802.6). Attempts to fit the same model as \citet{Carpano2018} to the continuum did not result in an acceptable fit, with a $\chi_r^2$=1.17 for 515 dof ($\chi^2$=604.7), unphysical and unconstrained parameters and clear structure in the residuals, particularly for the 2001 and 2003 spectra. Untying the parameters, does little to improve the fit nor does deleting components to emulate the models of \citet{Fornasini2017}. The \emph{TBabs*pcfabs(po*highecut)} model has an identical $\chi_r^2$=1.17 for 515 dof ($\chi^2$=604.7), whereas the \emph{TBabs*(bbody+po*highecut)} had $\chi_r^2$ values of 1.39 ($\chi^2$=724.64) for 521 dof and 1.37 for 517 dof ($\chi^2$=709.19) for when the blackbody temperature components are tied across all observations or allowed to vary respectively.

The continuum model of \citet{Bozzo2011} and \citet{VanDerMeer2005}, \emph{TBabs(po+TBabs(po))} with the two power law components constrained to be equal within an observation but allowed to vary between observations, produces a much better fit to the continuum - a $\chi_r^2$ of 1.07 ($\chi_r$ 558.6) for 541 dof. The $n_{H,i}$ values derived from the \emph{XMM-Newton} data vary from $\sim6\times10^{22}\text{ to } 2\times10^{24}$~cm$^{-2}$, whilst the photon indices range from a $\sim1-3$. For full details of the model fits and parameters see Table \ref{tab:xray_spec}. Whilst the fit is formally a good fit to the spectrum, we note that there are still some clear residuals in the soft excess of the spectra from 2003 and, to a lesser extent, the first spectrum of 2012. 

We also obtain a good fit to the spectrum when the first power law, representing the reflected emission, is replaced with a blackbody, \emph{TBabs(bbody+TBabs(po))} in \textsf{xspec}; $\chi_r^2$=1.05 ($\chi_r$= 542.1) for 515 dof. Similar values are obtained for the intrinsic absorption and power law components. This model seems to provide a better fit to the soft excess of the 2003 spectrum, but worse to the first spectrum of 2012. The parameters of this model are also included in Table \ref{tab:xray_spec}.
 
For completeness we also attempt to fit the continuum with the model presented by \citet{Ueda1998} fit to \emph{ASCA} data of CI~Cam taken during outburst. They fit the 0.8--10.0~keV spectrum with a two temperature Raymond-Smith \citep{Raymond1977} model (emission from a hot, diffuse gas) with an additional narrow Gaussian to better characterise the iron line region. We again ignore the 6.2--7.0 keV energy range to focus only on the continuum and fit a \emph{tbabs*tbabs*(vraymond$_1$+vraymond$_2$)} to the data, with the temperatures, abundances and flux ratios of the two components fixed at the values derived by \citeauthor{Ueda1998} (1998, i.e. the normalisation of $vraymond_1$ is $\sim4$ times that of $vraymond_2$). This leads to a  $\chi_r^2$=2.87 for 521 dof ($\chi^2$=1494.7). Removing the constraint that the flux ratio must stay equal to that reported in \citet{Ueda1998} does little to improve the fit with an identical $\chi_r^2$ value ($\chi_r^2$= 2.87 for 516 dof; $\chi^2$=1483.1). Allowing the temperatures to vary, as well as the fluxes, improves the fit ($\chi_r^2$= 1.23 for 514 dof; $\chi^2$=633.8), however, despite not being tied the temperature values for one of the components is pegged at the hard limit, suggesting that this model is not an accurate description of the physical processes occurring in the system. This is discussed further in Sect. \ref{sect:disc_outburst_v_quiescence}.

\subsubsection{Fe Emission lines}

\begin{figure}
  \centering
  \includegraphics[width=0.49\textwidth]{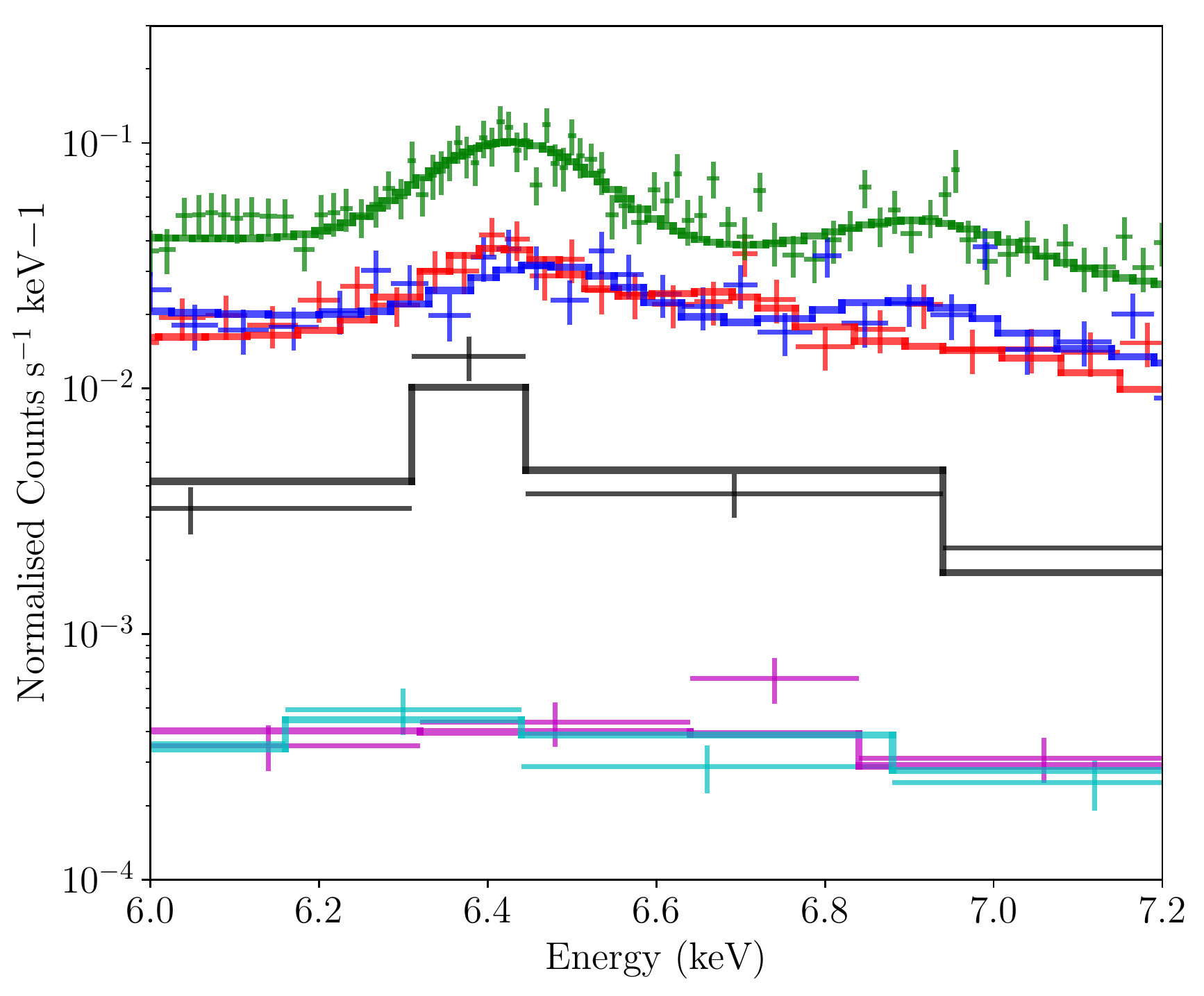}
  \caption{A close up of the 6.0--7.2 keV energy range that includes the region initially excluded from the spectral fits, along with the best fit \emph{TBabs(bbody+TBabs(po)+Gaus+Gaus)} model (though we note that both models give identical parameters for the iron lines within errors). The \emph{NuSTAR} data has been divided by 10 in this plot for clarity. The colour coding is the same as for Fig. \ref{fig:spectra}}\label{fig:Fe}
\end{figure}

The iron line region of the spectrum is shown in Figure \ref{fig:Fe}. When the iron line region is included in the spectrum, the fits understandably deteriorate (by $\Delta\chi_r^2\sim$+0.6). The addition of a single, intrinsic narrow Gaussian to the spectrum, constrained to be at the same energy across all observations but with the normalisation allowed to vary, results in the $\chi_r^2$ values falling to 1.17 and 1.16 for the \emph{TBabs(po+Tbabs(po))} and  \emph{TBabs(bbody+Tbabs(po))} models respectively. The resulting fits place the line at  6.43$\pm0.01$ keV, suggesting near neutral iron and consistent with that reported by \citet{Bartlett2013}. These values fall further to $\chi_r^2$=1.11/1.14 (\emph{po/bbody}) a second, constrained Gaussian is included. This is well constrained at 6.90$_{-0.04}^{+0.03}$~keV for the \emph{TBabs(po+Tbabs(po))} model, but poorly constrained for the \emph{TBabs(bbody+Tbabs(po))} model, 5.8$_{-0.4}^{+1.3}$~keV. Both fits improve greatly when the energies of these lines, as well as the normalisations, are allowed to vary, suggesting the ionisation of the iron is changing with time. The final line energies and normalisations (i.e. line fluxes) are included in Table \ref{tab:xray_spec}. 

\subsection{X-ray Timing}\label{sect:xray_timing}

\begin{figure*}
\centering
  \includegraphics[width=0.99\textwidth]{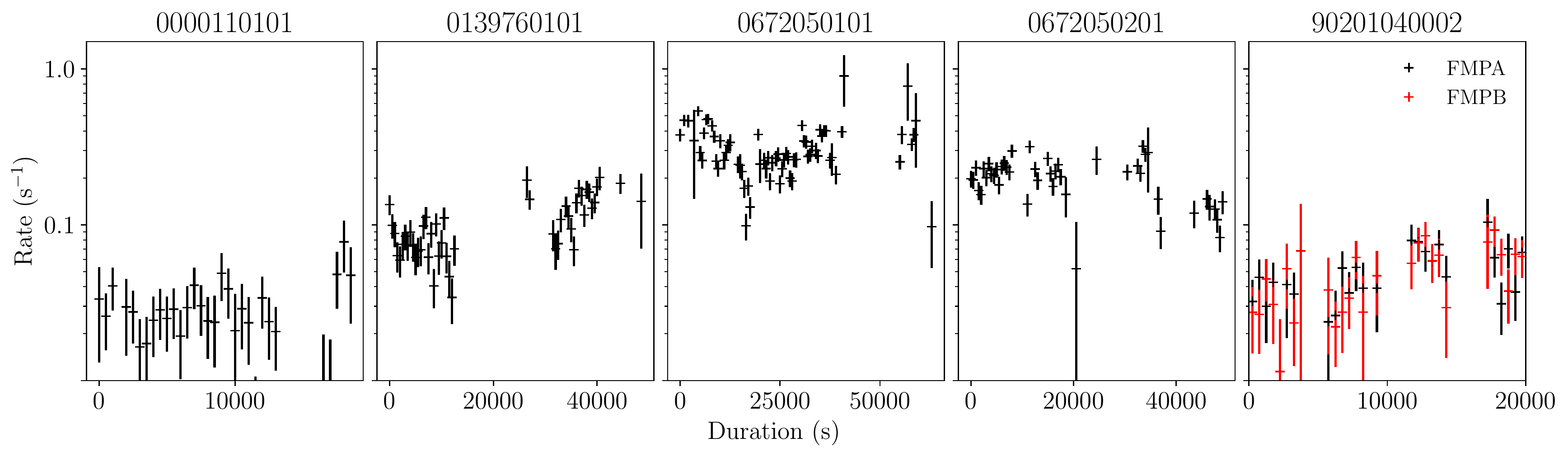}
  \caption{Background subtracted, 500~s bin light curves of CI Cam for each of the 4 \emph{XMM-Newton} EPIC-pn observations and the \emph{NuSTAR} observation. The \emph{XMM-Newton} light curves cover the 0.2--12.0~keV energy range, while the \emph{NuSTAR} light curve covers the 3.0-79.0~keV range. All observations are shown on the same logarithmic scale to facilitate easy comparison.}\label{fig:xmm_nustar_curve}
  \end{figure*}

Figure \ref{fig:xmm_nustar_curve} shows the background subtracted light curves of CI Cam for the \emph{XMM-Newton} and \emph{NuSTAR} observations. Table  
 \begin{table}
 \centering
 \caption{Properties of the X-ray light curves of CI Cam}\label{tab:xmm_lc}
 \begin{tabular}{lcc}
 \hline\hline\noalign{\smallskip}
 	& Rate & $F_{var}$ \\ \noalign{\smallskip}
\hline\noalign{\smallskip}
2001-08-19  &  0.029 $\pm$0.002  &  0.2$\pm$0.2  \\\noalign{\smallskip}
2003-02-24  &  0.104 $\pm$0.003 &  0.36$\pm$0.01  \\\noalign{\smallskip}
2012-02-20  &  0.318 $\pm$0.009  &  0.34$\pm$0.01  \\\noalign{\smallskip} 
2012-02-22  &  0.200 $\pm$0.005  &  0.26$\pm$ 0.01\\\noalign{\smallskip} 
2016-10-20:FMPA  &  0.057$\pm$0.004  &  0.21$\pm$ 0.07\\\noalign{\smallskip}
2016-10-20:FMPB  &  0.059$\pm$0.003  &  0.15$\pm$ 0.09\\
\hline
\end{tabular}
\end{table} \ref{tab:xmm_lc} gives the X-ray properties of the light curves in Fig. \ref{fig:xmm_nustar_curve}, including the fractional root mean square variability amplitude, $F_{var}$, calculated from equations (10) and (B3) of \citet{Vaughan2003}. This value includes the effects of both intrinsic source variability and flux measurement errors. It's clear that the source can vary significantly even within an observation, but this appears to be uncorrelated with the flux value itself as the values of $F_{var}$ for the observation in 2003 and the first observation in 2012 are consistent within errors, despite the factor $\sim$3 increase in the count rate. However, we must be cautious when drawing conclusions from these data; the spectra are complex with a variable soft component and absorbing column. Rather than comparing the $F_{var}$ to the total flux a better comparison could be the normalisation of the photon index that represents the intrinsic emission of the accretor (i.e. the second power law in the \emph{TBabs(po+TBabs(po)+Gaus+Gaus)} model, see Tab. \ref{tab:xray_spec}). Indeed, these values paint a very different picture, with the normalisations of the intrinsic power law components consistent (within errors) for the 2003 spectrum and the first spectrum of 2012, despite the factor 3 increase in the total count rate. 

The 0.1~s bin \emph{XMM-Newton} light curves were searched for pulsations using the \textsf{astropy.stats.LombScargle} routine \citep{Lomb1976,Scargle1982,Astropy2013,Astropy2018}. The Lomb-Scargle periodogram, is a commonly used tool for detecting periodic or quasi-periodic behaviour in different types of astrophysical objects. A heuristic was used to determine a suitable frequency grid for each observation, which in practise lead to a linear range of frequencies probing periods from approximately the bin size ($\sim0.1$~s) of the light curve to $\sim\times10$ the observation durations ($100$s of ks) with a step size of $\sim1.5\times10^{-6}$~s$^{-1}$. No periods were detected in any of the light curves. The \emph{NuSTAR} data were not searched for periods as the combination of low count rate and the window function due to the low earth orbit of the satellite make an independent detection of a period unlikely.

Figure \begin{figure}
  \centering
  \includegraphics[width=0.49\textwidth]{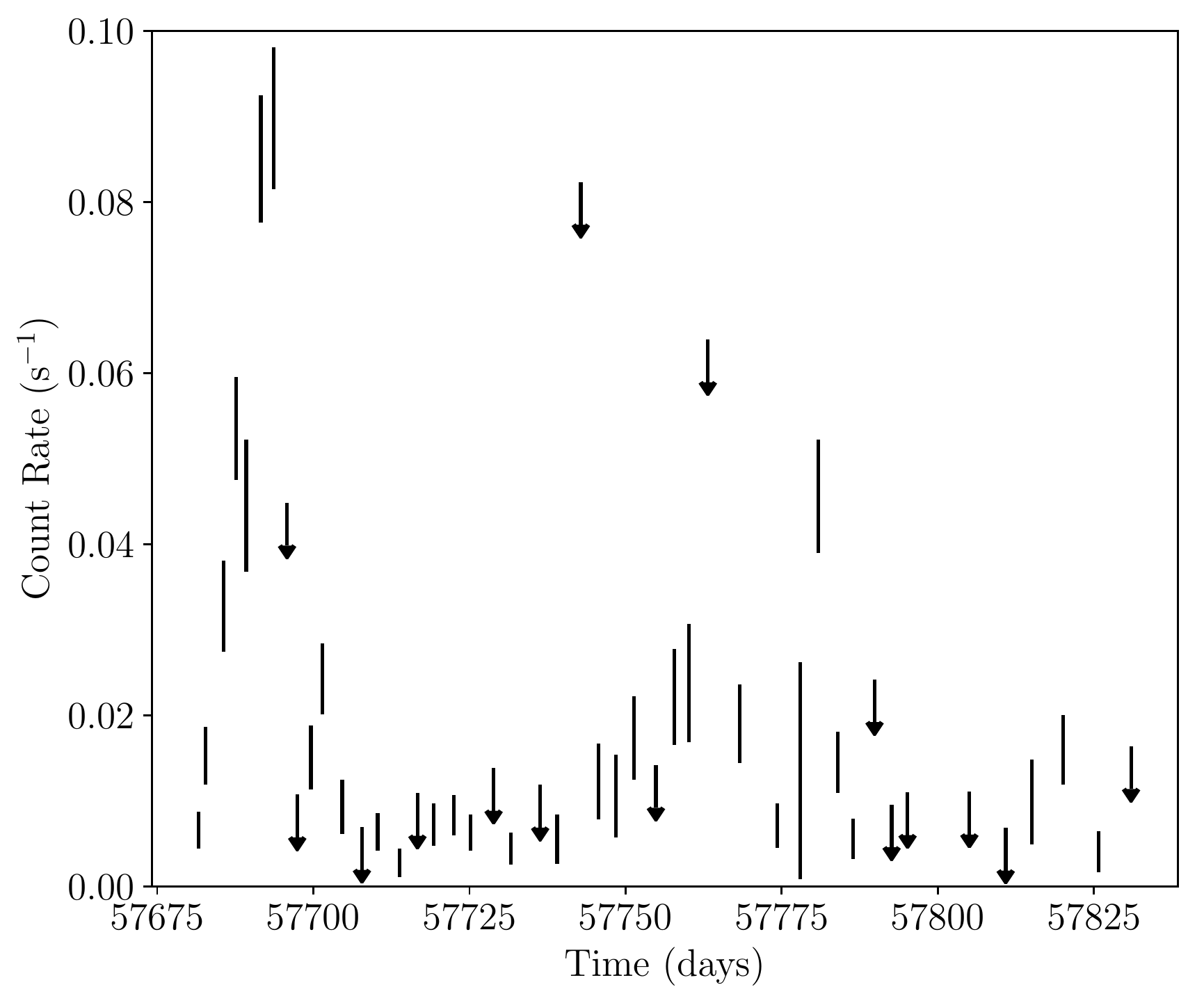}
  \caption{\emph{Swift} light curve spanning 150~days of snapshot observations of CI Cam.}\label{fig:sw_curve}
\end{figure}\ref{fig:sw_curve} shows the 0.3--10.0 keV light curve of the \emph{Swift} observations. The details of the each observation are included in Table \ref{tab:swift}. The light curve probes a different timescale of variability to those from the individual \emph{XMM-Newton} or \emph{NuSTAR} observations: on the order of days to months, rather than hours. The $F_{var}$ for the 150~day period spanned is 1.31$\pm$0.02. No evidence is found for the 19.4~day period of \citet{Barsukova2006}. 

\section{Discussion}

\subsection{Observational Findings}\label{sect:disc_obs}

Our optical spectrum is fully consistent with an explanation that CI~Cam was observed in a quiescent phase. The line profiles clearly point to a very complex, aspherical circumstellar envelope, which cannot be easily interpreted given a single snapshot. The extreme variability of the He\,{\sc ii} 4686{\AA} line leads us to treat the 19.4~day periodicity claimed by \citet{Barsukova2006} with caution; the transient nature of this line strongly suggests that it is produced in the stellar wind, rather than in the star itself, making it unsuitable for tracing dynamics. The same conclusion has been reached for other massive binary systems, including the HMXB/micro quasar SS\,433; \citet{Goranskij2011, Medvedev2013} and \citet{Robinson2017} all concluded that the He\,{\sc ii} 4686{\AA} line in this system originates from the disk wind and not close to the star or the compact object with \citet{Robinson2017} explicitly noting against using the He\,{\sc ii} lines as a proxy for the orbital radial velocity. It has also been shown that the He\,{\sc ii}~4686\AA{} line in the Wolf-Rayet HMXB IC\,10 X-1 \citep{Clark2004} does not directly trace the motion of the donor star \citep{Laycock2015}. 

The X-ray spectra can be interpreted as arising from emission from an accreting compact object (i.e. power law emission) with a reprocessed component that dominates the continuum at energies $<3$~keV and is the source of the iron lines (represented either by a second power law or a blackbody). This is produced in a dense circumstellar environment, which also contributes a substantial obscuring column (with an additional contribution from the interstellar medium.) The source of the reprocessed emission appears to be outside that of the absorbing column. It is unclear how such a system can occur but this is likely due to the geometry of the system. We note that our X-ray spectral fits are likely an approximation of the true situation, but the low count rates in our data preclude further analysis.

\begin{figure}
\centering
\includegraphics[width=0.49\textwidth]{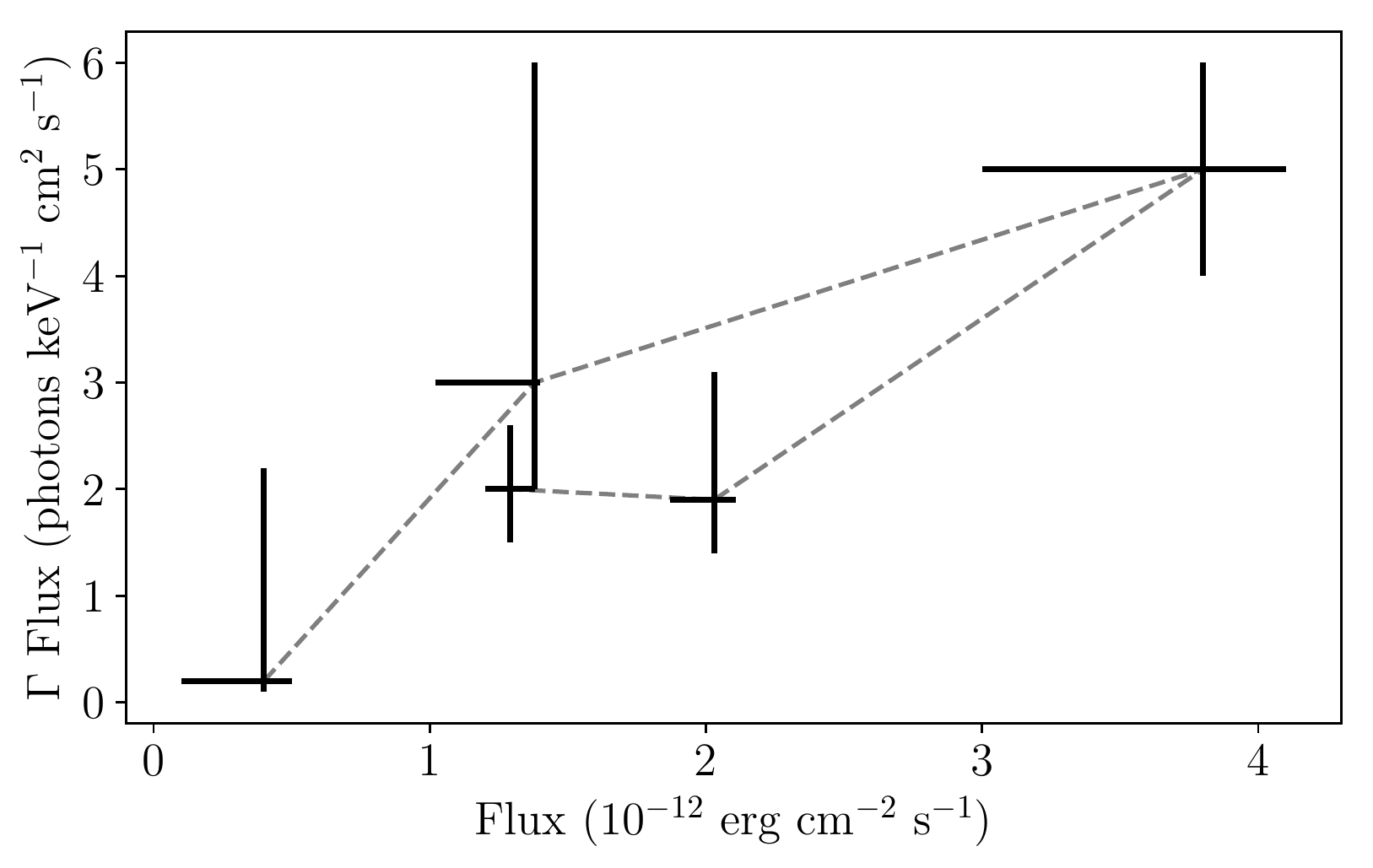}\\
\includegraphics[width=0.49\textwidth]{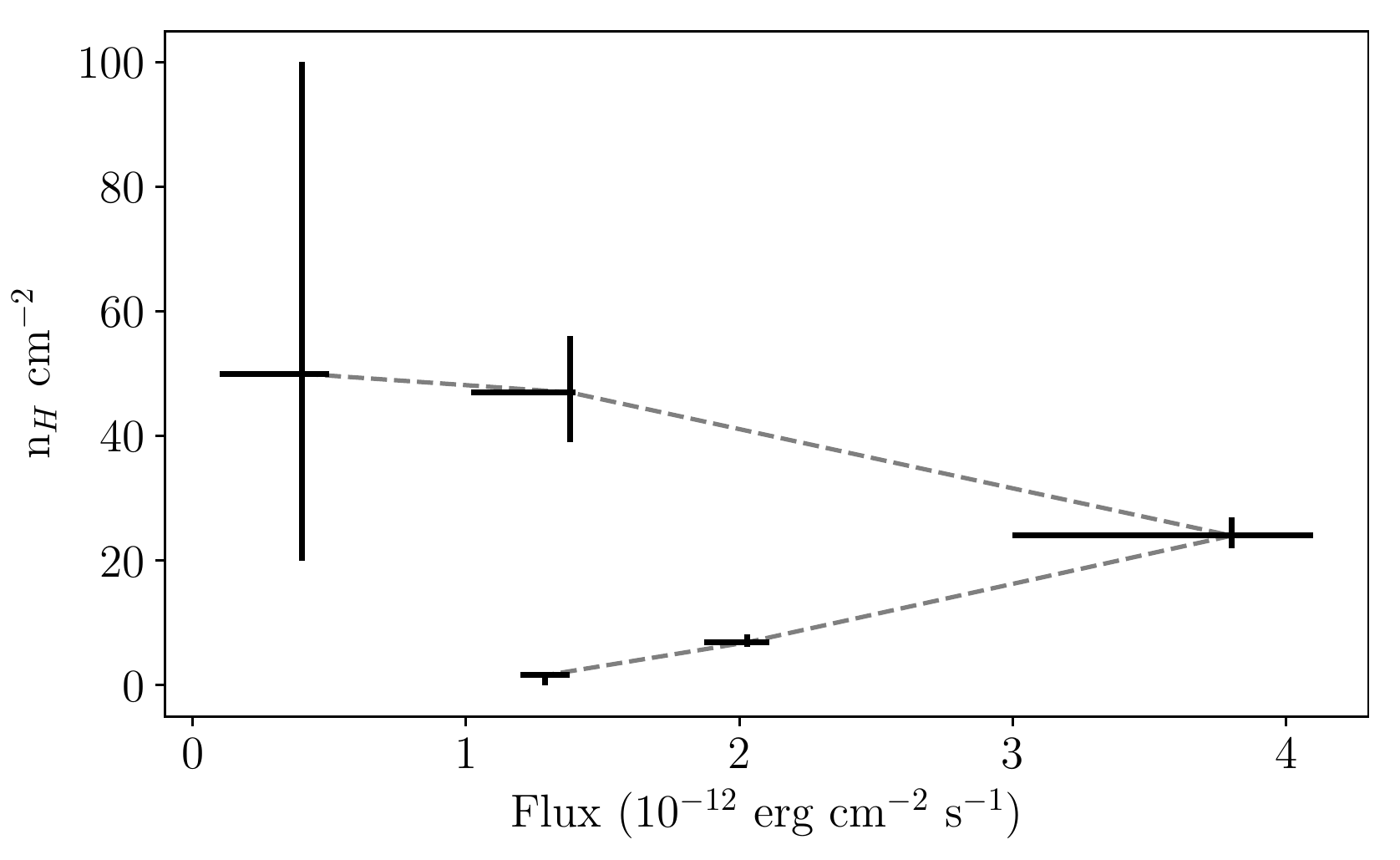}\\
\includegraphics[width=0.49\textwidth]{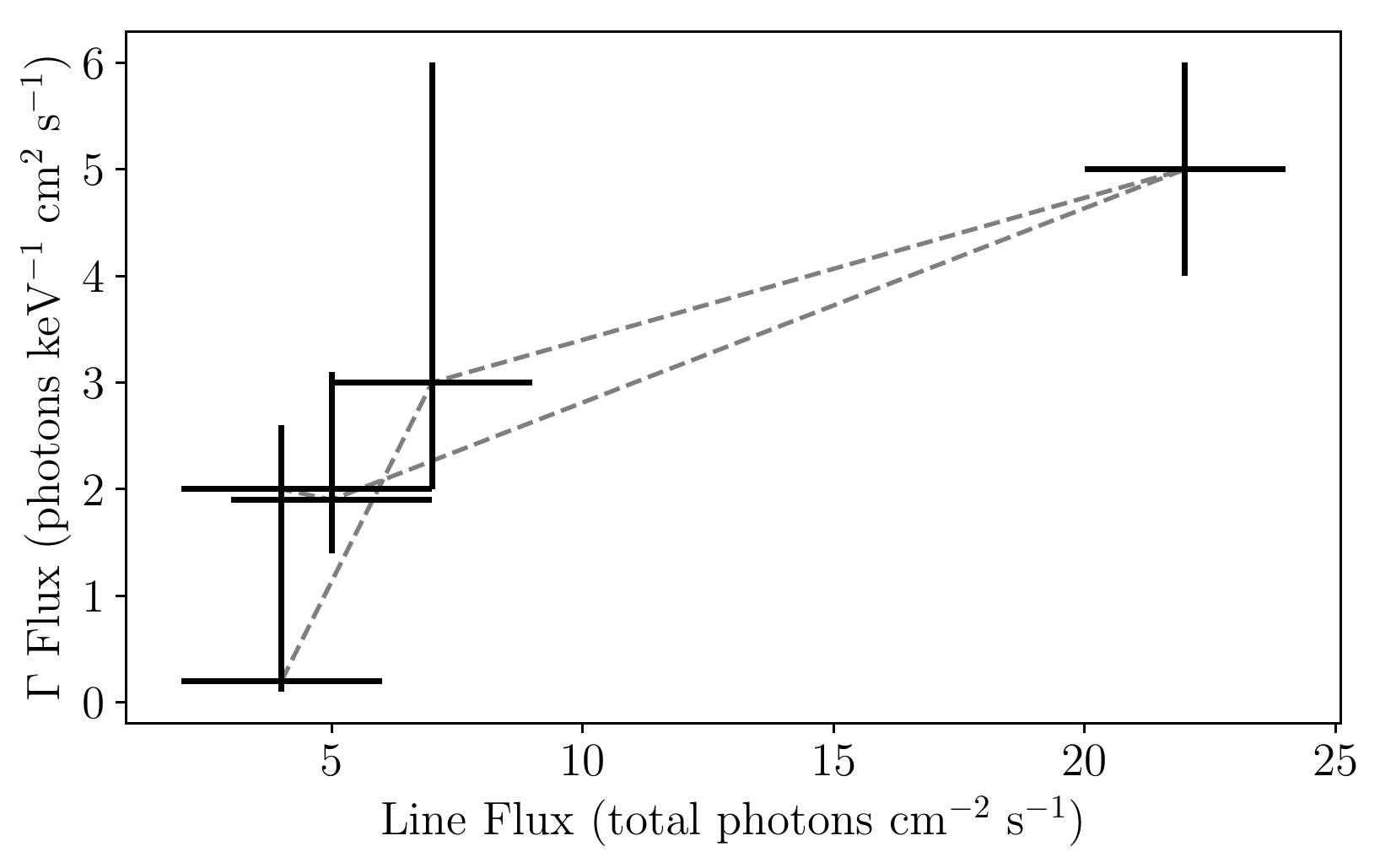}\\
\includegraphics[width=0.49\textwidth]{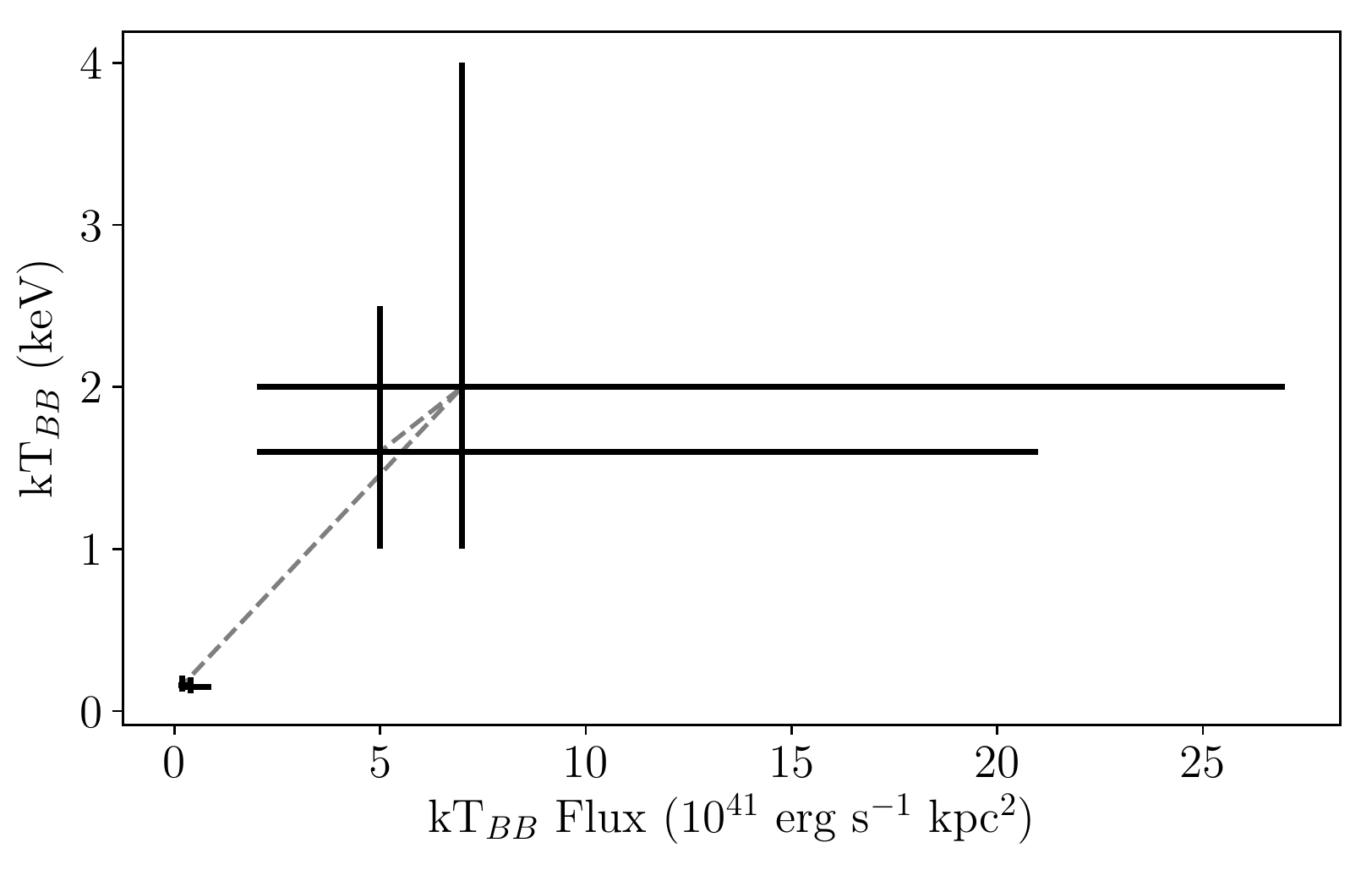}\\
\caption{Comparison of the parameters \emph{TBabs*(bbody+TBabs(powerlaw)+Gaussian+Gaussian)}  model fit to the data, discussed in the text. Energy$_1$ is the energy of the first Fe-K$\alpha$ line.}\label{fig:correlations}
\end{figure}

Despite being in quiescence, CI Cam is highly X-ray variable, on timescales of days, in terms of both integrated flux and spectral shape - although the same physical components are present in each spectrum. Figure \ref{fig:correlations} shows how the different X-ray model components of the \emph{TBabs(bbody+TBabs(po)+Gaus+Gaus)} model vary with respect to each other. Some of the parameters are poorly constrained making it difficult to draw meaningful conclusions; however it is clear from both Fig \ref{fig:spectra} and  \ref{fig:correlations} that there is a stronger correlation between the normalisation (i.e. flux) of the power law component and the total flux than between the column density and total flux. This strongly implies that the increase in observed flux is due to an intrinsic increase of the total flux of the system (i.e. soft excess plus intrinsic emission), rather than a decrease in the absorbing column. This conclusion is reinforced by the correlation seen between the power law flux and line flux: with more photons available for reprocessing, the Fe-K$\alpha$ line becomes stronger. The second iron line in the model is not as well constrained but it is clear from Table \ref{tab:xray_spec} that the energies of both lines are varying, however, the relationship between the energies of these two lines is not obvious. The strongest correlation between any of the parameters is found between the characteristic temperature and flux of the soft excess - the soft excess becomes simultaneously hotter and brighter.

We attribute the changes in the X-ray spectra of CI Cam to change in the environment local to the compact object: Changes in the circumstellar environment due to clumping and/or asymmetries lead to changes in the line of sight absorbing column to the X-ray source and variations in the amount of material local to the compact object available for accretion. This in turn could be due to varying mass loss and/or system geometry or, alternatively, due to orbital effects as the compact object moves through its orbit. Such drastic changes between the two 2012 spectra would argue for a short, eccentric orbit (i.e. not inconsistent with that proposed by \citealt{Barsukova2006}) but with no clear orbital period available, it is not possible to distinguish between these two scenarios. The \emph{Swift} data seem to suggest variability on timescales of $\sim75-100$~days, but further monitoring with a similar cadence is required before this could be further interpreted. We note, however, that \citet{Iyer2017} recently discovered an $\sim$80~day period in the 12~year X-ray light curve of IGR~J16318-4848, a source that shares many similarities with CI Cam (see Sect \ref{sect:disc_comp})

We see no evidence for X-ray pulsations, although given the flux level this is not unexpected. The absence of pulsations during the 1998 outburst is, however, unexpected. Nevertheless, systems such as the supergiant XRB IGR J18214-1318 \citep{Fornasini2017} also lack pulsations despite having an X-ray spectrum strongly indicative of a neutron star and indeed similar to that of CI Cam. In the absence of pulsations we cannot unambiguously determine the nature of the accretor.

\subsection{The Distance to CI Cam}

Another shortcoming is that, even with the two data releases from \emph{Gaia}, we still cannot improve on previous estimates on the distance to CI Cam. \citet{Wijngaarden2016} claim a distance of 1.4~kpc based on a simple inversion of the \emph{Gaia} DR1 parallax (0.7$\pm$0.2). However, this is explicitly warned against by the \emph{Gaia} team themselves \citep{Luri2018}. Furthermore, the \emph{Gaia} DR2 parallax value, 0.09$\pm$0.03 is inconsistent with that from DR1. Figure \begin{figure}
  \centering
  \includegraphics[width=0.49\textwidth]{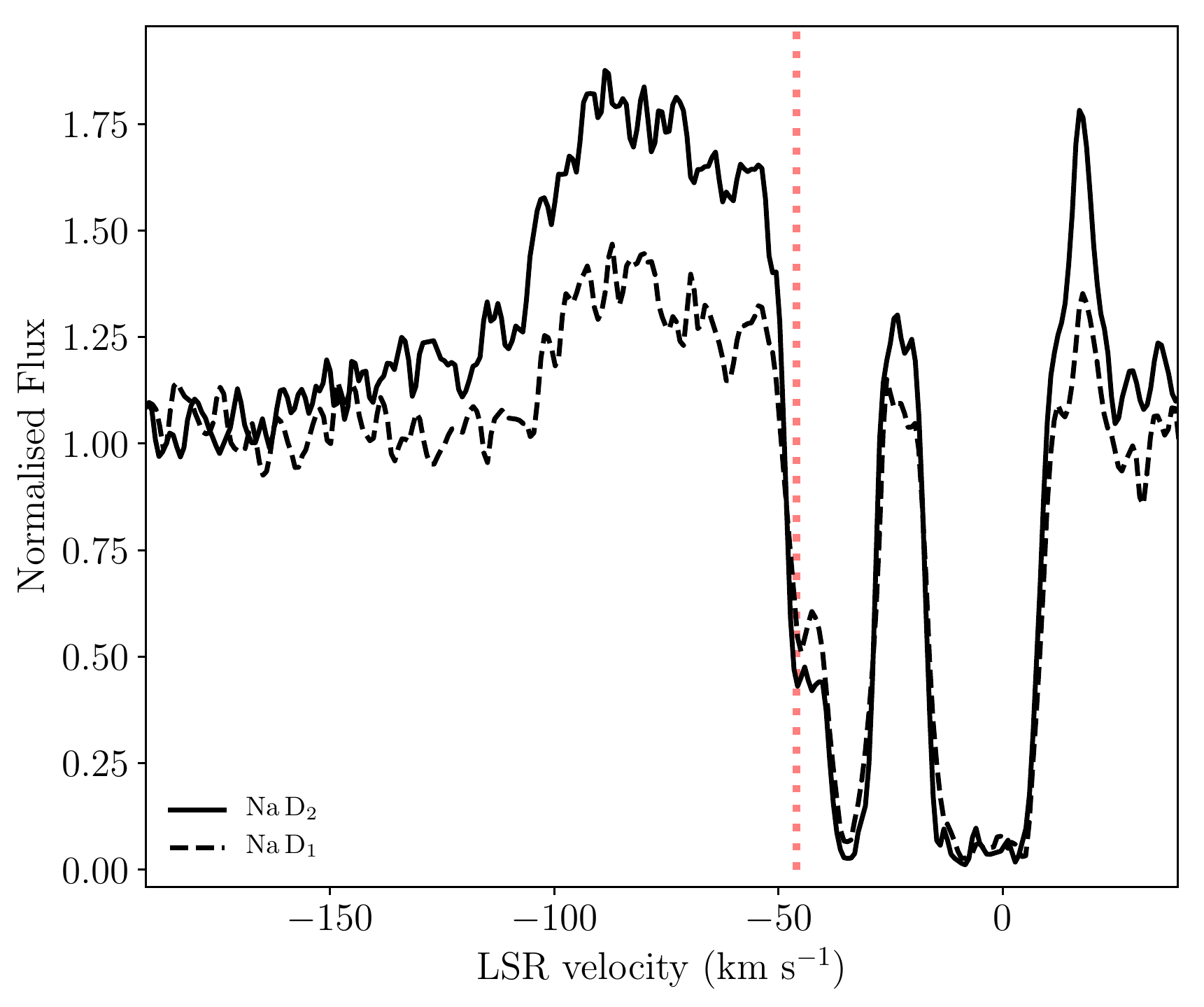}
  \caption{The velocities of the interstellar Na\,{\sc D} absorption lines in the Local Standard of Rest.}\label{fig:NaD}
\end{figure}\ref{fig:NaD} shows the interstellar Na\,{\sc D} lines present in our spectrum of CI Cam, in the Local Standard of Rest (LSR). \citet{Hynes2002} argue for a possible lower limit of 6-8~kpc on the distance to CI Cam, based on a weak absorption feature at $\sim-45$~km~s$^{-1}$. Our higher resolution spectra clearly resolve this feature at around -46~km~s$^{-1}$. This value is very similar to the systemic velocity of the system (51$\pm2$~km~s$^{-1}$, \citealt{Aret2016}), suggesting that the radial velocity of CI Cam could be dominated by Galactic dynamics. If we assume this is the case and calculate the \citet{Reid2014} Galactic rotation curve in the direction of CI~Cam, a LSR velocity of -46~km~s$^{-1}$ corresponds to a kinematic distance estimate of $\sim5.5$~kpc; the presence of this feature in the spectrum of CI~Cam, suggests that the source is more distant than this value.

The distance to CI Cam is important since this uncertainty accommodates a wide range of possible quiescent ($L_{0.2-12 {\rm keV}}\sim 1.0-10.0\times10^{33}(d/5 {\rm kpc})^2$~ergs~s$^{-1}$; This work) and outburst X-ray luminosities ($L_{2-25 {\rm keV}}\sim 3.0\times10^{38}(d/5 {\rm kpc})^2$~ergs~s$^{-1}$; \citealt{Robinson2002}; distances $\gtrsim$ than 2~kpc lead to an outburst flux greater than the Eddington limit of a white dwarf, whilst at large distances the outburst approaches the Eddington limit for a neutron star. The distance uncertainty also has implications for the integrated panchromatic outburst luminosity and the quiescent luminosity of the mass donor (from which one might hope to infer a mass via comparison to isochrones). 

\subsection{Comparison of outburst and quiescent data}\label{sect:disc_outburst_v_quiescence}

The presence of X-ray emission twenty years after the outburst clearly shows that the accretor is bound to the mass donor - one cannot easily appeal to a transitory interaction with an isolated compact object (cf. \citealt{Hynes2002}). Given the presence of  all-sky monitoring missions, we can be relatively certain that no comparable event to the 1998 outburst has occurred since. Four post-outburst observations between 1998-2001 indicate a comparable level of quiescence emission\footnote{For a distance of 5~kpc $L_{1-10{\rm keV}} \sim 1.4\times10^{33}$~ergs~s$^{-1}$ to 
$2.3\times10^{34}$~ergs~s$^{-1}$ with a single non-detection of $<1.4\times10^{33}$~ergs~s$^{-1}$  in 2000 \citep{Orlandini2000,Parmar2000,Boirin2002}} to that which we report here. 

Intriguingly the basic form of the X-ray emission (soft and hard components with substantial local extinction; e.g. \citealt{Orr1998, Belloni1999, Boirin2002}) and the reasons for variability (changes in emission \emph{and} obscuration) in quiescent and outburst appears broadly comparable. For example the spectrum of CI Cam during the short-lived soft X-ray flaring episodes that occurred during the decay phase of the 1998 outburst \citep{Ueda1998} can be qualitatively understood as short lived periods of reduced circumstellar extinction. Figure\begin{figure}
  \centering
  \includegraphics[width=0.49\textwidth]{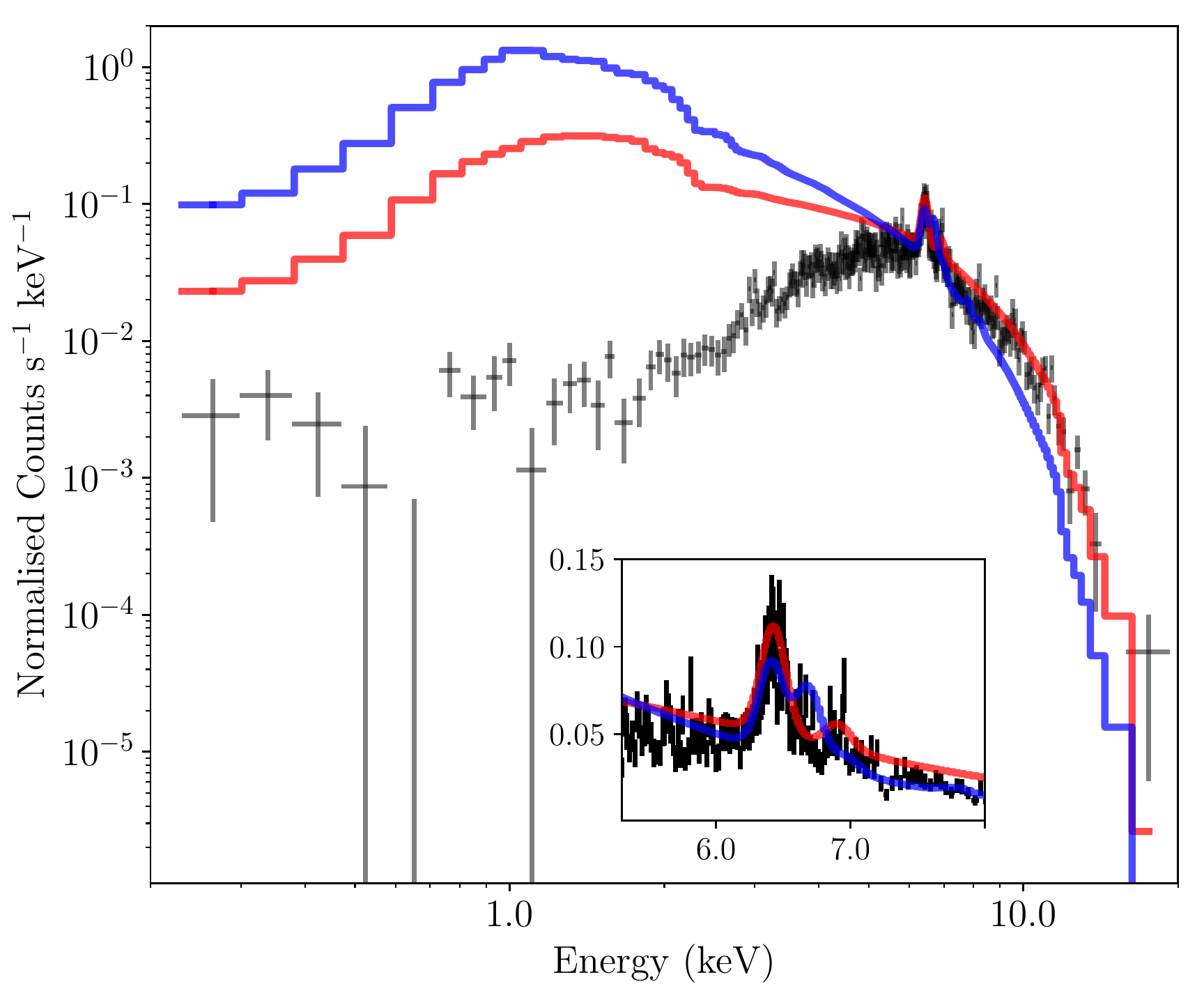}
  \caption{The third \emph{XMM-Newton} spectrum of CI Cam with the best fit \emph{TBabs(TBabs(bbody+po)+Gaus+Gaus)} model in red, with the intrinsic absorption (i.e. the second \emph{TBabs} term) set to zero along with the best fit model to the outburst data from \citet{Ueda1998} in blue. The normalisation of the outburst model has been adjusted for easier comparison.}\label{fig:Ueda}
\end{figure} \ref{fig:Ueda} shows the third of the four \emph{XMM-Newton} spectra (the spectrum with the greatest number of counts) with the best fit \emph{TBabs(TBabs(bbody+po)+Gaus+Gaus)} model with the intrinsic absorption ($n_{H,i}$) set to zero. The data has also been fit with the best fit model of \citealt{Ueda1998}, with only the normalisation of the first temperature component allowed to vary (the normalisation of the second hot gas component was fixed to be 0.25 times that of the first, as in \citealt{Ueda1998}) and the normalisation of the additional Gaussian free to vary. The shape of the two spectra are similar, though not identical. \citet{Belloni1999} similarly reported highly variable extinction in the \emph{RXTE} data taken during outburst.

\subsection{Comparison to other systems}\label{sect:disc_comp}

The sgB[e] star IGR J16318-4848 \citep{Filliatre2004} is another apparently persistent X-ray binary that shows rapid flaring and variable local extinction \citep{Ibarra2007, Courvoisier2003, Krimm2010}. As with CI Cam the distance is highly uncertain, as is the nature of the accretor, for which no pulsations have been identified \citep{Barragan2009}. Assuming a slightly closer distance to IGR J16318-4848 (1.6-4 kpc; \citealt{Chaty2012}) the quiescent luminosities of this source and CI Cam are very similar. Both sources also have the same characteristic spectra with high levels of absorption ($\times10^{23-24}$~cm$^{-2}$) and strong iron line complex. However, IGR~J16318-4848 does not appear to have the same level of variation in its column density or iron line in quiescence \citep{Ibarra2007}. Additionally IGR J16318-4848 doesn't appear to have undergone an event comparable to the 1998 outburst of CI Cam, instead it has undergone two smaller outbursts, separated by $\sim9$ years with peak luminosities $L_X\sim10^{36}$~ergs~s$^{-1}$ assuming a distance of 4~kpc \citep{Filliatre2004}. As noted in Sect \ref{sect:disc_obs}, there has been a recent, tentative claim of an $\sim80$~day period in this target: similar to the timescales of variability suggested by our \emph{Swift} observations.

 Another object of interest is S18. While the limited X-ray observations suggest that it is less luminous than IGR J16318-4848 - and indeed consistent with a colliding wind binary - the lower interstellar extinction towards it allows access to the He\,{\sc ii} 4686{\AA} transition which is found to be highly variable, with no obvious comparator amongst either single massive stars or CWBs.

The most compelling comparator, however, is the intermediate luminosity optical transient/supernova imposter SN2009da (=NGC\,300 ULX-1; see below). Supernovae imposters such as SN2000ch (=NGC 3432 2000-OT; \citealt{Pastorello2010}) are objects which experience (multiple) rapid photometric outbursts akin to low luminosity SNe but with the crucial difference that the progenitor star typically survives such events. However an increasing number of such events are found to presage the loss of such stars to core-collapse events shortly thereafter\footnote{SN2006jc \citep{Pastorello2007}, SN2010mc \citep{Ofek2013} and potentially SN2009ip \citealt{Mauerhan2013}, but see \citealt{Fraser2013}.}. As a consequence the nature and physical cause of such behaviour is of considerable interest, with many authors highlighting the similarities between SNe imposters and the giant eruptions of massive luminous blue variables (LBVs) such as the 19th century event in $\eta$~Carina. 

First detected at optical wavelengths \citep{Monard2010}, SN2010da is of interest in the context of CI Cam for a number of reasons. Firstly photometric and spectroscopy strongly suggest that the primary is a moderately luminous ($\sim10^4L_{\odot}$) sgB[e] star \citep{Lau2016,Villar2016}; indeed the spectrum of SN2010da  bears particularly close similarity to that of CI Cam (Sect. \ref{sect:xray_spec}). Secondly the optical outburst was accompanied by an X-ray flare with a flux of $6\times10^{38}$~ergs~s$^{-1}$ \citep{Immler2010}, significantly in excess of the Eddington luminosity of a neutron star. Subsequent  observations over the following four years revealed SN2010da to be a persistent, variable X-ray source and hence a likely HMXB \citep{Binder2011, Binder2016}. This  hypothesis was dramatically confirmed by the detection of pulsations by \citet{Carpano2018} indicative of a neutron star, while detailed modelling of the X-ray spectrum showed close similarities to that of CI Cam; comprising a power law and reprocessed component subject to substantial local extinction.

The similarity of CI Cam to SN2010da across mid-IR, optical and X-ray wavelengths\footnote{Subject to distance determination for CI Cam to determine X-ray and bolometric luminosities.} is particularly striking, suggesting that if CI Cam was located in an external galaxy its 1998 outburst would have led to a classification as a supernova imposter. As such the proximity to CI Cam allows us to investigate one channel leading to supernovae imposters/intermediate luminosity optical transients in unprecedented detail. \citet{Hynes2002} suggest that one possible origin for the 1998 event in CI Cam was the interaction of a shock wave generated by impulsive accretion onto the compact object with the dense surrounding circumstellar envelope, with \citet{Lau2016} suggesting an identical scenario for SN2010da, in a manner analogous to type IIn supernovae. Critically \citet{Mioduszewski2004} provide multiple-epoch radio continuum observations of the expansion of just such a shock front through the circumstellar environment of CI Cam. The highly structured radio emission provides compelling evidence that the circumstellar envelope of CI Cam is likewise asymmetric and clumpy. In such a picture X-ray variability can arise from both changes in the accretion rate onto the compact object and variability in the line of sight extinction, with the latter mechanism clearly operating in the quiescent X-ray observations considered here. Given these similarities we consider it likely that CI Cam likewise contains a neutron star, but this scenario is ultimately independent of this assumption.

However problems still remain. Most obviously we still lack an ultimate explanation for the 1998 and 2010 outbursts of CI Cam and SN2010da. Specifically, it is not clear what caused the X-ray outbursts which appear to initiate the optical-radio flares. Possibilities include accretion from a highly dynamic circumstellar envelope e.g. a transient episode of enhanced mass loss from the primary impacted upon the accretor, although then one has to explain the physical mechanism that led to this event. Alternatively one might suppose a very wide and eccentric orbit for the compact object, such that the enhanced accretion occurred at periastron. Given the apparent lack of similar events in CI Cam this would suggest a minimum orbital period of 20~years. 


\section{Concluding Remarks}

In this paper we present a detailed analysis of the quiescent X-ray properties of CI Cam, the archetypal sgB[e]/X-ray binary, motivated by reports of a potential new outburst in the system. We present new and both published and unpublished archival data from \emph{Swift}, \emph{XMM-Newton} and \emph{NuSTAR} as well as a high resolution MERCATOR/HERMES optical spectrum.

Despite being in quiescence, we find that CI Cam is highly X-ray variable, on timescales of days, both in terms of total integrated flux and spectral shape. All the spectra can be described by a heavily absorbed power law along with a soft excess at energies $<3$~keV, represented by either a second power law or blackbody component in the spectra, and iron line region. This iron line region can be further decomposed into two, intrinsically narrow iron lines: A near neutral Fe-K$\alpha$ line and a highly ionised Fe-K$\alpha$ line. All the model components vary by a factor of $\gtrsim$10 though there is no obvious correlation between most of the parameters.

Qualitatively, many of the variation in the X-ray properties of CI Cam can be explained by the presence of an accreting compact companion (black hole, neutron star or white dwarf) immersed in a dense, highly structured, aspherical circumstellar envelope. The differences in the accretion flux and circumstellar extinction could represent either changes in this environment triggered by variable mass loss from the star, or the conditions local to the accretor by virtue of orbital motion (or indeed a combination of the two). The exact nature of the compact object remains unclear. The photon index of the X-ray spectrum is consistent with those reported for accreting neutron stars, however, a thorough search for X-ray pulsations, indicative of an accreting pulsar (i.e. a neutron star) did not return any periods. Some of this uncertainty will be resolved when definitive distance to CI Cam is determined, likely in a later \emph{Gaia} data release.

The outbursts from SNe imposters often precede core-collapse events, but the cases of CI Cam and SN2010da seem to highlight an alternative scenario in which the star/system can survive. CI Cam is particularly interesting in this context as its proximity allows us to observe this behaviour in unprecedented detail. This includes directly imaging the progression of the shock front through the circumstellar envelope, analogous to the interaction of the shockwaves in type Ib/IIN SNe interacting with pre-existing ejecta.

\begin{acknowledgements}
      We would to thank the anonymous referee whose comments improved the quality of this manuscript. We would also like to thank Willem-Jan de Wit, for an interesting discussion on discs vs rings and the [e] phenomenon; Sergio Sim{\' o}n Di\'{a}z, for providing us with the HERMES observations CI Cam and Javier Lorenzo, for calculating the Galactic rotation curve in the direction of CI Cam. This research was supported by the UK Science and Technology Facilities Council and the Spanish Ministerio de Ciencia, Innovaci\'{o}n y Universidades under grant AYA2015-68012-C2-2-P (MINECO/FEDER). This work has made use of data from the European Space Agency (ESA) mission \emph{Gaia} (\url{https://www.cosmos.esa.int/gaia}), processed by the \emph{Gaia} Data Processing and Analysis Consortium (DPAC,
\url{https://www.cosmos.esa.int/web/gaia/dpac/consortium}). Funding for the DPAC has been provided by national institutions, in particular the institutions participating in the \emph{Gaia} Multilateral Agreement. This research is based on observations obtained with XMM-Newton, an ESA science mission with instruments and contributions directly funded by ESA Member States and NASA. Based on observations made with the Mercator Telescope, operated on the island of La Palma by the Flemmish Community, at the Spanish Observatorio del Roque de los Muchachos of the Instituto de Astrof\'{i}sica de Canarias. Based on observations obtained with the HERMES spectrograph, which is supported by the Research Foundation - Flanders (FWO), Belgium, the Research Council of KU Leuven, Belgium, the Fonds National de la Recherche Scientifique (F.R.S.-FNRS), Belgium, the Royal Observatory of Belgium, the Observatoire de Gen\`{e}ve, Switzerland and the Th\"{u}ringer Landessternwarte Tautenburg, Germany. This work made use of data from the \emph{NuSTAR} mission, a project led by the California Institute of Technology, managed by theJet Propulsion Laboratory, and funded by the National Aeronautics and Space Administration. This work made use of data supplied by the UK Swift Science Data Centre at the University of Leicester. We would like to thank the \emph{NuSTAR} PI, Fiona Harrison, and the late Neil Gehrels, the PI of \emph{Swift}, for accepting our observations in Director's Discretionary Time.
 
 \end{acknowledgements}


\bibliographystyle{aa}
\bibliography{cicam}

\begin{appendix}
\onecolumn
\section{Log of the \emph{Swift} Observations included in this study}
\begin{table*}[h!]
 \centering
 \caption{Log of the \emph{Swift}-XRT monitoring observations. All errors reported are at the 1$\sigma$ level, all upper limits are all 3$\sigma$ upper limits.}\label{tab:swift}
 \begin{tabular}{lcccclccc}
 \hline\hline\noalign{\smallskip}
 Obs. ID		& Obs. Date	& Exp. Time	& Count Rate 		&		 &  Obs. ID		& Obs. Date	& Exp. Time	& Count Rate \\\noalign{\smallskip}
 			&			& (s)			& counts s$^{-1}$ 	&		 &  				&			& (s)			& counts s$^{-1}$ \\ \noalign{\smallskip}
\hline\noalign{\smallskip}
00031511002  &  2016-10-16  & 994.9            	& 0.027$_{-0.008}^{+0.010}$ 	& & 00031511026 	&  2016-12-17  & 1024.6             & 0.006$_{-0.002}^{+0.003}$ \\\noalign{\smallskip}
00031511003  &  2016-10-20  & 1980.4          	& 0.007$\pm0.002$			& & 00031511027  &  2016-12-20  & 82.2             	& $<0.08$ \\\noalign{\smallskip}
00031511005  &  2016-10-21  & 1848.8           	& 0.015$\pm0.003$			& & 00031511028  &  2016-12-23  & 995.4            	& 0.013$_{-0.004}^{+0.005}$ \\\noalign{\smallskip}
00031511006  &  2016-10-24  & 1801.3           	& 0.033$\pm0.005$ 			& & 00031511029  &  2016-12-26  & 722.3             	& 0.011$_{-0.004}^{+0.005}$ \\\noalign{\smallskip}
00031511007  &  2016-10-26  & 1962.8            	& 0.054$\pm0.006$			& & 00031511030  &  2016-12-29  & 923.2            	& 0.018$_{-0.004}^{+0.005}$ \\\noalign{\smallskip}
00031511008  &  2016-10-28  & 979.3          	& 0.044$\pm0.008$			& & 00031511031  &  2017-01-01  & 966.9             	& $<0.01$ \\\noalign{\smallskip}
00031511009  &  2016-10-30  & 2059.6          	& 0.085$\pm0.007$			& & 00031511032  &  2017-01-04  & 870.9             	& 0.022$\pm0.006$ \\\noalign{\smallskip}
00031511010  &  2016-11-01  & 1641.9             & 0.090$\pm0.008$ 			& & 00031511033  &  2017-01-07  &  825.2            	& 0.024$_{-0.006}^{+0.008}$ \\\noalign{\smallskip}
00031511011  &  2016-11-03  & 539.1            	& $<0.04$ 				& & 00031511034  &  2017-01-10  &  363.0            	& $<0.06$ \\\noalign{\smallskip}				
00031511012  &  2016-11-05  & 1149.2            	& $<0.01$ 				& & 00031511036  &  2017-01-15  & 1214.6           	& 0.019$\pm0.005$\\\noalign{\smallskip}		
00031511013  &  2016-11-07  & 1645.7            	& 0.015$\pm0.004$			& & 00031511037  &  2017-01-21  & 1474.3          	& 0.007$_{-0.002}^{+0.003}$ \\\noalign{\smallskip}	 
00031511014  &  2016-11-09  & 1816.0             & 0.024$\pm0.004$ 			& & 00031511038  &  2017-01-24  & 185.5             	& 0.017$_{-0.010}^{+0.016}$ \\\noalign{\smallskip}
00031511015  &  2016-11-12  & 1220.7            & 0.01$_{-0.003}^{+0.004}$ 	& & 00031511039  &  2017-01-27  & 1290.4             & 0.046$\pm0.007$ \\\noalign{\smallskip}
00031511016  &  2016-11-15  & 1956.0           	& $<0.007$ 				& & 00031511040  &  2017-01-30  & 1394.2            	& 0.014$\pm0.004$\\\noalign{\smallskip}
00031511017  &  2016-11-18  & 1856.4           	& 0.007$\pm0.002$			& & 00031511041  &  2017-02-02  & 1494.4             & 0.006$_{-0.002}^{+0.003}$ \\\noalign{\smallskip}
00031511018  &  2016-11-21  & 1918.8             & 0.003$_{-0.001}^{+0.002}$ 	& & 00031511042  &  2017-02-05  &  647.1            	& <0.02 \\\noalign{\smallskip}
00031511019  &  2016-11-24  & 1160.9             & $<0.01$ 				& & 00031511043  &  2017-02-08  & 1262.8             & <0.01 \\\noalign{\smallskip}			
00031511020  &  2016-11-27  & 1574.9             & 0.007$_{-0.002}^{+0.003}$ 	& & 00031511044  &  2017-02-11  & 1165.5             	& <0.01 \\\noalign{\smallskip}
00031511021  &  2016-11-30  & 2014.3             & 0.009$_{-0.002}^{+0.003}$ 	& & 00031511045  &  2017-02-21  & 1218.2            	& <0.01 \\\noalign{\smallskip}
00031511022  &  2016-12-03  & 2011.5             & 0.006$\pm0.002$ 			& & 00031511046  &  2017-02-26  & 1653.7            	& <0.007 \\\noalign{\smallskip}
00031511023  &  2016-12-06  & 1352.5             & $<0.01$ 				& & 00031511047  &  2017-03-03  & 556.1              	& 0.011$_{-0.004}^{+0.006}$ \\\noalign{\smallskip}			
00031511024  &  2016-12-09  &1868.0            	& 0.005$\pm0.002$ 			& & 00031511048  &  2017-03-08  & 1222.5             & 0.016$\pm0.004$ \\\noalign{\smallskip}
00031511025  &  2016-12-14  & 932.2             	& $<0.01$ 				& & 00031511049  &  2017-03-13  & 1308.8             & 0.004$_{-0.002}^{+0.003}$ \\\noalign{\smallskip} 
		       &		      &				& 						& & 00031511050  &  2017-03-18  & 1484.3            	& <0.02 \\\noalign{\smallskip}
\hline
\end{tabular}
\end{table*}
\newpage
\begin{landscape}
\section{Best Fit Model Parameters}
\begin{table}[h!]
\centering
\begin{threeparttable}
\caption{Table of the best fit model parameters to spectra of CI Cam. All errors stated are the 90\% confidence limit unless otherwise stated.}\label{tab:xray_spec}
\begin{tabular}{lcccccccccccc}
\hline\hline\noalign{\smallskip}
 & \multicolumn{11}{c}{Model 1: \emph{TBabs(po+TBabs(po)+Gaus+Gaus)}; $\chi^2_r$ = 1.10}\\
\hline\noalign{\smallskip}

		& 			& norm $\Gamma_{refl}$\tnote{(a)}	& $n_{H,i}$	 		& $\Gamma$		& norm $\Gamma_{i}$	& $E_{line,1}$			& norm $E_{line,1}$\tnote{(b)}  & $E_{line,2}$			& norm $E_{line,2}$ &  $C_{FPMB}$ & \multicolumn{2}{c}{Flux$_{obs}$\tnote{(c)}} \\\noalign{\smallskip}
		
	 	&			& $10^{-6}$		& ($10^{22}$ cm$^{-2}$)	&				& $10^{-4}$			& (keV)				& $10^{-6}$		& (keV)				& $10^{-6}$		&		& \multicolumn{2}{c}{10$^{-12}$ ergs~cm$^{-2}$~s$^{-1}$} \\\noalign{\smallskip}
\hline\noalign{\smallskip}
2001-08-19		&			& 6$\pm$3			& 130$\pm40$			& 2.8$_{-0.7}^{+1.0}$& 65$_{-61}^{+290}$	& 6.38$\pm0.05$		& 3$\pm$2		&  -					&  -			&					& \multicolumn{2}{c}{0.31$_{-0.27}^{+0.04}$}	\\\noalign{\smallskip}

2003-02-24		&			& 2.7$\pm$0.9			& 54$_{-8}^{+9}$		& 1.2$\pm$0.3		& 4$_{-2}^{+4}$			& 6.41$\pm0.03$		& 7$\pm$2		& 6.67$\pm0.06$		& 3$_{-2}^{+1}$	&					& \multicolumn{2}{c}{1.4$_{-0.28}^{+0.06}$}	\\\noalign{\smallskip}

2012-02-20  	&			& 7$\pm$1			& 20$\pm$2			& 1.0$\pm$0.1		& 4$\pm$1			& 6.43$\pm0.01$		& 22$\pm$2		& 6.90$\pm0.04$		& 6$\pm$2	&					& \multicolumn{2}{c}{3.82$_{-0.13}^{+0.06}$} \\\noalign{\smallskip}

2012-02-22	&			& 9$\pm$3			& 6.3$\pm0.8$			& 1.2$\pm$0.1		& 2.5$_{-0.5}^{+0.7}$	& 6.49$_{-0.07}^{+0.06}$	& 5$\pm$2		& 6.90$_{-0.12}^{+0.09}$	& 3$\pm$2	&					& \multicolumn{2}{c}{2.04$\pm0.06$} \\\noalign{\smallskip}

2016-10-20	&			& -\tnote{(d)}			& $<1.6$				& 1.6$_{-0.2}^{+0.1}$& 2.3$_{-0.5}^{+0.6}$	& 6.3$_{-0.7}^{+0.2}$	& 4$\pm$3		& 6.8$\pm0.2$\tnote{(e)}	& $<4$		& 1.1$\pm0.1$			& 1.3$_{-0.2}^{+0.1}$  & 4.7$_{-0.6}^{+0.5}$\\\noalign{\smallskip}

\hline\noalign{\smallskip}
 & \multicolumn{11}{c}{Model 2: \emph{TBabs(bbody+TBabs(po)+Gaus+Gaus)}; $\chi^2_r$ = 1.07}\\
\hline\noalign{\smallskip}
		& $kT_{BB}$ 	& norm $kT_{BB}$\tnote{(f)}		& $n_{H,i}$	 		& $\Gamma$		& norm $\Gamma_{i}$	& $E_{line,1}$							& norm $E_{line,1}$ & $E_{line,2}$			& norm $E_{line,2}$ & $C_{FPMB}$ & \multicolumn{2}{c}{Flux$_{obs}$}	\\ \noalign{\smallskip}
		
	 	&			& $10^{-6}$			& ($10^{22}$ cm$^{-2}$)	&				& $10^{-4}$			& (keV)								& $10^{-6}$		& (keV)				& $10^{-6}$ &		&	  \multicolumn{2}{c}{10$^{-12}$ ergs~cm$^{-2}$~s$^{-1}$} \\\noalign{\smallskip}
\hline\noalign{\smallskip}

2001-08-19		&	0.15$_{-0.04}^{+0.06}$	& 0.4$_{-0.2}^{+0.5}$	& 50$_{-30}^{+50}$	& 0.4$_{-0.6}^{+1.3}$	& 0.2$_{-0.1}^{+2.0}$	& 6.38$_{-0.05}^{+0.06}$		& $4\pm2$		&	-  					&	-		& 	& \multicolumn{2}{c}{0.4$_{-0.3}^{+0.1}$} 	 \\\noalign{\smallskip}

2003-02-24		&	0.16$_{-0.04}^{+0.06}$	& 0.2$_{-0.1}^{+0.2}$	& 47$_{-8}^{+9}$	& 1.0$\pm0.3$			& 3$_{-1}^{+3}$			& 6.41$_{-0.03}^{+0.02}$		& 7$\pm2$		& 6.68$_{-0.06}^{+0.05}$		& 3$\pm1$	&  	&\multicolumn{2}{c}{1.38$_{-0.36}^{+0.02}$} \\\noalign{\smallskip}

2012-02-20	&	2$_{-1}^{+2}$			& 7$_{-5}^{+20}$		& 24$_{-2}^{+3}$	& 1.1$_{-0.2}^{+0.3}$	& 5$\pm1$			& 6.431$_{-0.008}^{+0.011}$	& 22$\pm2$ 		& 6.90$\pm0.05$			& 5.8$\pm0.2$	&	&  \multicolumn{2}{c}{3.8$_{-0.8}^{+0.3}$}  	\\\noalign{\smallskip}

2012-02-22	& 1.6$_{-0.6}^{+0.9}$		& 5$_{-3}^{+16}$		&6.9$_{-0.7}^{+1.3}$ & 1.1$_{-0.2}^{+0.6}$	& 1.9$_{-0.5}^{ +1.2}$	& 6.48$_{-0.05}^{+0.06}$		& 5$\pm2$		&  6.89$_{-0.09}^{+0.08}$		&3$\pm2$ 	&	&  \multicolumn{2}{c}{2.03$_{-0.16}^{+0.08}$}  \\\noalign{\smallskip}

2016-10-20	& -						&	-\tnote{(d)}			&	$<1.6$		&  1.5$\pm0.1$			& 2.0$_{-0.5}^{+0.6}$	&  6.3$_{-0.7}^{+0.2}$		& 4$_{-2}^{+3}$		&  6.8$_{-0.2}^{+0.1}$\tnote{(e)}	& $<5$		&1.1$\pm0.1$	&  1.29$\pm0.09$	& 4.7$_{-0.4}^{+0.6}$ \\\noalign{\smallskip}
\hline	
\end{tabular}
\begin{tablenotes}\footnotesize
\item[(a)] Normalisation of power law components in units of photons keV$^{-1}$~cm$^{-2}$~s$^{-1}$ at 1 keV.
\item[(b)] Normalisation of Gaussian components in units of total photons~cm$^{-2}$~s$^{-1}$ in the line.
\item[(c)] Flux values calculated over 0.2-10.0 keV for the \emph{XMM-Newton} spectra and 2.0-10.0 and 2.0-79.0~keV respectively for the \emph{NuSTAR} spectra.
\item[(d)] \emph{NuSTAR} response does not extend below 3~keV where this component dominates in the \emph{XMM-Newton} spectra and so the normalisation of this component is fixed at zero.
\item[(e)] 1$\sigma$ error; \textsf{xspec} could not compute the 90\% confidence limit.
\item[(f)] Normalisation of Blackbody component in units of $L_{39}/D_{10}^2$, where $L_{39}$ is the luminosity of the Blackbody in units of $10^{39}$ ergs~s$^{-1}$ and $D_{10}$ is the distance to the source in units of 10~kpc.
\end{tablenotes}
\end{threeparttable}
\end{table}
\end{landscape}

\end{appendix}

 \end{document}